%% file: main.tex
\pdfoutput=1

\documentclass[12pt,a4paper]{article}

\usepackage{ifthen} 
\newboolean{pdflatex}
\setboolean{pdflatex}{true} 

\usepackage{overpic}

\newboolean{articletitles}
\setboolean{articletitles}{true} 

\newboolean{uprightparticles}
\setboolean{uprightparticles}{false} 

\newboolean{inbibliography}
\setboolean{inbibliography}{false} 

\input{preamble}
\usepackage{longtable} 

\begin{document}

\renewcommand{\thefootnote}{\fnsymbol{footnote}}
\setcounter{footnote}{1}

\input{title-LHCb-PAPER}


\renewcommand{\thefootnote}{\arabic{footnote}}
\setcounter{footnote}{0}



\pagestyle{plain} 
\setcounter{page}{1}
\pagenumbering{arabic}


%

\input{paper-introduction}
\input{paper-detector}
\input{paper-selection}
\input{paper-efficiency}
\input{paper-fit}
\input{paper-systematics}
\input{paper-results}

\input{acknowledgements}

\addcontentsline{toc}{section}{References}
\setboolean{inbibliography}{true}
\bibliographystyle{LHCb}
\bibliography{main,LHCb-PAPER,LHCb-CONF,LHCb-DP}

\end{document}

%% file: preamble.tex
\usepackage{microtype}
\usepackage{lineno}  
\usepackage{xspace} 

\usepackage{graphicx}  
\usepackage{color}
\usepackage{colortbl}
\graphicspath{{./figs/}} 

\usepackage{amsmath} 
\usepackage{amssymb}
\usepackage{amsfonts}
\usepackage{upgreek} 

\newcommand*\patchAmsMathEnvironmentForLineno[1]{%
\expandafter\let\csname old#1\expandafter\endcsname\csname #1\endcsname
\expandafter\let\csname oldend#1\expandafter\endcsname\csname
end#1\endcsname
 \renewenvironment{#1}%
   {\linenomath\csname old#1\endcsname}%
   {\csname oldend#1\endcsname\endlinenomath}%
}
\newcommand*\patchBothAmsMathEnvironmentsForLineno[1]{%
  \patchAmsMathEnvironmentForLineno{#1}%
  \patchAmsMathEnvironmentForLineno{#1*}%
}
\AtBeginDocument{%
\patchBothAmsMathEnvironmentsForLineno{equation}%
\patchBothAmsMathEnvironmentsForLineno{align}%
\patchBothAmsMathEnvironmentsForLineno{flalign}%
\patchBothAmsMathEnvironmentsForLineno{alignat}%
\patchBothAmsMathEnvironmentsForLineno{gather}%
\patchBothAmsMathEnvironmentsForLineno{multline}%
}

\usepackage{hyperref}    
\usepackage[all]{hypcap} 

\input{lhcb-symbols-def} 

\usepackage{cite} 
\usepackage{mciteplus}

%% file: lhcb-symbols-def.tex
\def\lhcb {\mbox{LHCb}\xspace}

\ifthenelse{\boolean{uprightparticles}}%
{

 \def\Ppsi        {\ensuremath{\uppsi}\xspace}

 \def\PDelta      {\ensuremath{\Delta}\xspace}                 
 \def\PXi      {\ensuremath{\Xi}\xspace}                 
 \def\PLambda      {\ensuremath{\Lambda}\xspace}                 
 \def\PSigma      {\ensuremath{\Sigma}\xspace}                 
 \def\POmega      {\ensuremath{\Omega}\xspace}                 
 \def\PUpsilon      {\ensuremath{\Upsilon}\xspace}                 
 
 \def\PB      {\ensuremath{\mathrm{B}}\xspace}                 
                  
 \def\PD      {\ensuremath{\mathrm{D}}\xspace}

 \def\PJ      {\ensuremath{\mathrm{J}}\xspace}                 
 \def\PK      {\ensuremath{\mathrm{K}}\xspace}

 \def\Pb      {\ensuremath{\mathrm{b}}\xspace}                 
 \def\Pc      {\ensuremath{\mathrm{c}}\xspace}

 \def\Pi      {\ensuremath{\mathrm{i}}\xspace}

 \def\Ps      {\ensuremath{\mathrm{s}}\xspace}

}
{

 \def\Ppsi        {\ensuremath{\psi}\xspace}                 
                  
 \mathchardef\PDelta="7101
 \mathchardef\PXi="7104
 \mathchardef\PLambda="7103
 \mathchardef\PSigma="7106
 \mathchardef\POmega="710A
 \mathchardef\PUpsilon="7107
                  
 \def\PB      {\ensuremath{B}\xspace}                 
                  
 \def\PD      {\ensuremath{D}\xspace}

 \def\PJ      {\ensuremath{J}\xspace}                 
 \def\PK      {\ensuremath{K}\xspace}

 \def\Pb      {\ensuremath{b}\xspace}                 
 \def\Pc      {\ensuremath{c}\xspace}

 \def\Pi      {\ensuremath{i}\xspace}

 \def\Ps      {\ensuremath{s}\xspace}

}

\def\squark    {\ensuremath{\Ps}\xspace}

\def\cquark    {\ensuremath{\Pc}\xspace}

\def\bquark    {\ensuremath{\Pb}\xspace}

\def\kaon  {\ensuremath{\PK}\xspace}
  \def\Kbar  {\kern 0.2em\overline{\kern -0.2em \PK}{}\xspace}

\def\Kp    {\ensuremath{\kaon^+}\xspace}
\def\Km    {\ensuremath{\kaon^-}\xspace}

\def\KS    {\ensuremath{\kaon^0_{\rm\scriptscriptstyle S}}\xspace} 
 
\def\Kstarz  {\ensuremath{\kaon^{*0}}\xspace}
\def\KstarzENT{\ensuremath{\kaon^{*}(892)^{0}}\xspace}

  \def\Dbar    {\kern 0.2em\overline{\kern -0.2em \PD}{}\xspace}

\def\B       {\ensuremath{\PB}\xspace}
\def\Bbar    {\ensuremath{\kern 0.18em\overline{\kern -0.18em \PB}{}}\xspace}

\def\Bu      {\ensuremath{\B^+}\xspace}

\def\Bd      {\ensuremath{\B^0}\xspace}
\def\Bs      {\ensuremath{\B^0_\squark}\xspace}

\def\jpsi     {\ensuremath{{\PJ\mskip -3mu/\mskip -2mu\Ppsi\mskip 2mu}}\xspace}

  \def\Y#1S{\ensuremath{\PUpsilon{(#1S)}}\xspace}

\def\Lz {\ensuremath{\PLambda}\xspace}
\def\Lbar {\ensuremath{\kern 0.1em\overline{\kern -0.1em\PLambda}}\xspace}

\def\Lb      {\ensuremath{\Lz^0_\bquark}\xspace}
\def\Lbbar   {\ensuremath{\Lbar^0_\bquark}\xspace}

\newcommand{\decay}[2]{\ensuremath{#1\!\to #2}\xspace}         

\def\to                 {\ensuremath{\rightarrow}\xspace}

\def\CP                {\ensuremath{C\!P}\xspace}
\def\CPT               {\ensuremath{C\!PT}\xspace}

\def\bToJPsiX     {\decay{H_b}{\jpsi X}}

\def\BsToJPsiPhi  {\decay{\Bs}{\jpsi\phi}}
\def\BdToJPsiKst  {\decay{\Bd}{\jpsi\Kstarz}}
\def\BdToJPsiKstENT{\decay{\Bd}{\jpsi\KstarzENT}}
\def\BdToJPsiKS   {\decay{\Bd}{\jpsi\KS}}
\def\BuToJPsiK    {\decay{\Bu}{\jpsi K^+}}

\def\LbToJPsiL    {\decay{\Lb}{\jpsi\Lz}}

\def\AT#1     {\ensuremath{A_{\mathrm{T}}^{#1}}\xspace}           

\def\C#1      {\ensuremath{\mathcal{C}_{#1}}\xspace}                       
\def\Cp#1     {\ensuremath{\mathcal{C}_{#1}^{'}}\xspace}                    
\def\Ceff#1   {\ensuremath{\mathcal{C}_{#1}^{\mathrm{(eff)}}}\xspace}        
\def\Cpeff#1  {\ensuremath{\mathcal{C}_{#1}^{'\mathrm{(eff)}}}\xspace}       
\def\Ope#1    {\ensuremath{\mathcal{O}_{#1}}\xspace}                       
\def\Opep#1   {\ensuremath{\mathcal{O}_{#1}^{'}}\xspace}                    

\newcommand{\tev}{\ifthenelse{\boolean{inbibliography}}{\ensuremath{~T\kern -0.05em eV}\xspace}{\ensuremath{\mathrm{\,Te\kern -0.1em V}}\xspace}}
\newcommand{\gev}{\ensuremath{\mathrm{\,Ge\kern -0.1em V}}\xspace}
\newcommand{\mev}{\ensuremath{\mathrm{\,Me\kern -0.1em V}}\xspace}
\newcommand{\kev}{\ensuremath{\mathrm{\,ke\kern -0.1em V}}\xspace}
\newcommand{\ev}{\ensuremath{\mathrm{\,e\kern -0.1em V}}\xspace}
\newcommand{\gevc}{\ensuremath{{\mathrm{\,Ge\kern -0.1em V\!/}c}}\xspace}
\newcommand{\mevc}{\ensuremath{{\mathrm{\,Me\kern -0.1em V\!/}c}}\xspace}
\newcommand{\gevcc}{\ensuremath{{\mathrm{\,Ge\kern -0.1em V\!/}c^2}}\xspace}
\newcommand{\gevgevcccc}{\ensuremath{{\mathrm{\,Ge\kern -0.1em V^2\!/}c^4}}\xspace}
\newcommand{\mevcc}{\ensuremath{{\mathrm{\,Me\kern -0.1em V\!/}c^2}}\xspace}

\def\mm   {\ensuremath{\rm \,mm}\xspace}

\def\mum  {\ensuremath{\,\upmu\rm m}\xspace}

\def\ps   {\ensuremath{{\rm \,ps}}\xspace}
\def\fs   {\ensuremath{\rm \,fs}\xspace}

\def\invps{\ensuremath{{\rm \,ps^{-1}}}\xspace}

\def\gsim{{~\raise.15em\hbox{$>$}\kern-.85em
          \lower.35em\hbox{$\sim$}~}\xspace}
\def\lsim{{~\raise.15em\hbox{$<$}\kern-.85em
          \lower.35em\hbox{$\sim$}~}\xspace}

\def\sPlot{\mbox{\em sPlot}}

\def\pt         {\mbox{$p_{\rm T}$}\xspace}

\def\evtgen     {\mbox{\textsc{EvtGen}}\xspace}

\def\geant      {\mbox{\textsc{Geant4}}\xspace}

\def\photos     {\mbox{\textsc{Photos}}\xspace}

\def\pythia     {\mbox{\textsc{Pythia}}\xspace}

\def\tell1  {TELL1\xspace}
\def\ukl1   {UKL1\xspace}

\newcommand{\DOCAz}{\mbox{$\rho$}\xspace}

%% file: title-LHCb-PAPER.tex

\begin{titlepage}
\pagenumbering{roman}

\vspace*{-1.5cm}
\centerline{\large EUROPEAN ORGANIZATION FOR NUCLEAR RESEARCH (CERN)}
\vspace*{1.5cm}
\hspace*{-0.5cm}
\begin{tabular*}{\linewidth}{lc@{\extracolsep{\fill}}r}
\ifthenelse{\boolean{pdflatex}}
{\vspace*{-2.7cm}\mbox{\!\!\!\includegraphics[width=.14\textwidth]{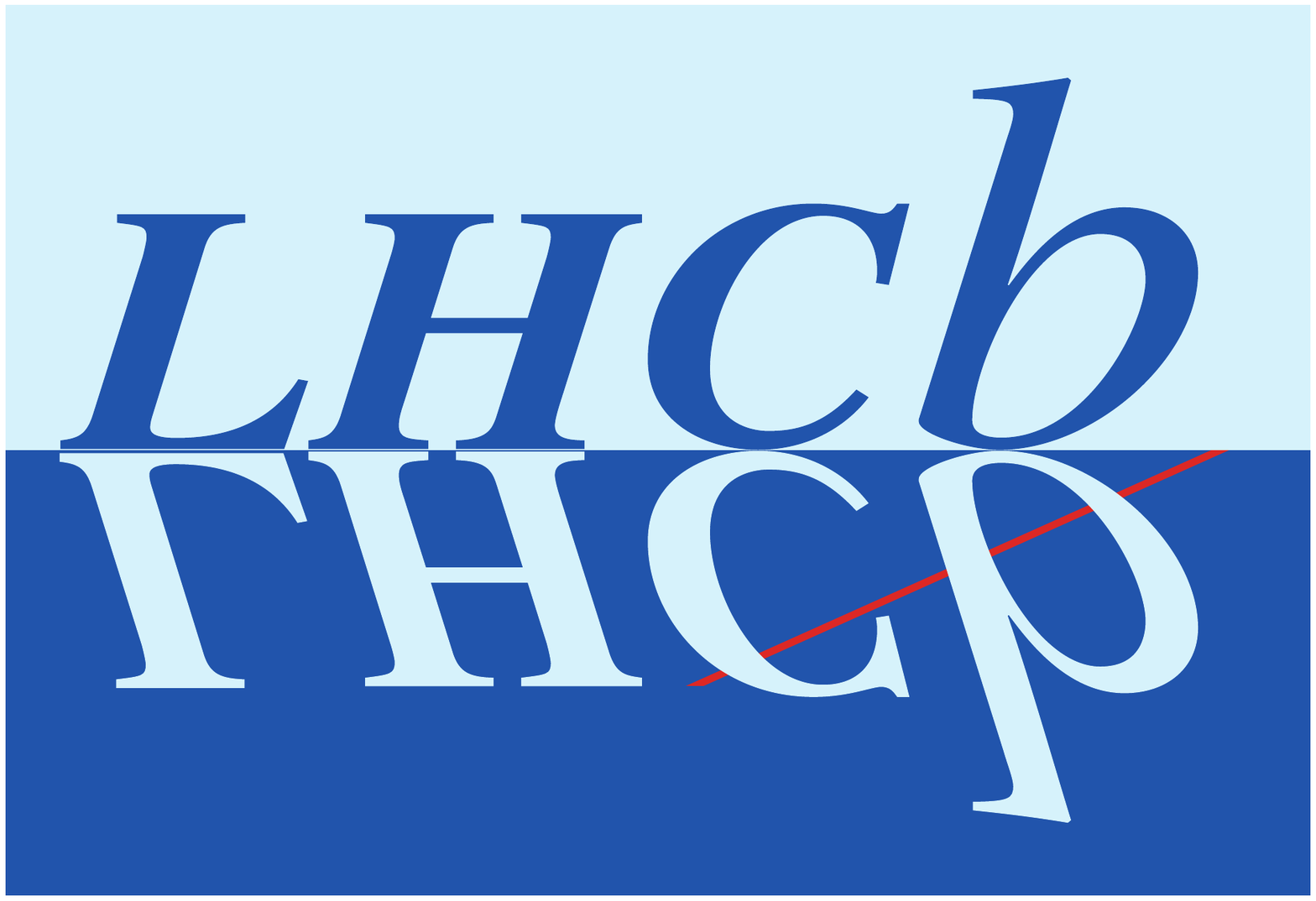}} & &}%
{\vspace*{-1.2cm}\mbox{\!\!\!\includegraphics[width=.12\textwidth]{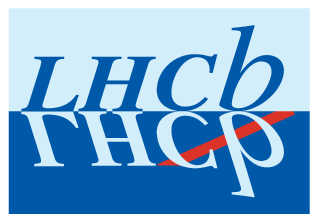}} & &}%
\\
 & & CERN-PH-EP-2014-018 \\  
 & & LHCb-PAPER-2013-065 \\  
 & & March 19, 2014 \\ 
 & & \\
\end{tabular*}

\vspace*{2.0cm}

{\bf\boldmath\huge
\begin{center}
  Measurements of the $\Bu$, $\Bd$, $\Bs$ meson and $\Lb$ baryon lifetimes
\end{center}
}

\vspace*{1.1cm}

\begin{center}
The LHCb collaboration\footnote{Authors are listed on the following pages.}
\end{center}

\vspace{\fill}

\begin{abstract}
  \noindent
  Measurements of $b$-hadron lifetimes are reported using $pp$
  collision data, corresponding to an integrated luminosity of 1.0\,fb$^{-1}$,
  collected by the LHCb detector at a centre-of-mass
  energy of $7$\tev. 
  Using the exclusive decays
  \BuToJPsiK, \BdToJPsiKstENT, \BdToJPsiKS, \LbToJPsiL and \BsToJPsiPhi
  the average decay times in these modes are measured to be
\begin{center}
    \begin{tabular}{lr}
      $\tau_{\BuToJPsiK}$		& =  	 $1.637\	\pm$ 0.004   $\pm$ 0.003  \ps,\\
      $\tau_{\BdToJPsiKst}$  	& =	 $1.524\	\pm$ 0.006   $\pm$ 0.004  \ps,\\
      $\tau_{\BdToJPsiKS}$  	& = 	 $1.499\	\pm$ 0.013   $\pm$ 0.005  \ps,\\
      $\tau_{\LbToJPsiL}$		&  =   $1.415\	\pm$ 0.027   $\pm$ 0.006  \ps,\\
      $\tau_{\BsToJPsiPhi}$		& = 	 $1.480\	\pm$ 0.011   $\pm$ 0.005  \ps,\\
      \end{tabular}
\end{center}
where the first uncertainty is statistical and the second is systematic.
These represent the most precise lifetime measurements in these decay modes.
In addition, ratios of these lifetimes, and the ratio of
the decay-width difference,
$\Delta\Gamma_d$, to the average width, $\Gamma_d$, in the $B^0$ system,
$\Delta \Gamma_d/\Gamma_d = -0.044 \pm 0.025 \pm 0.011$, are reported.
All quantities are found to be consistent with Standard Model expectations.
%
%
%
\end{abstract}

\vspace*{1.0cm}

\begin{center}
  Submitted to JHEP
\end{center}

\vspace{\fill}

{\footnotesize 
\centerline{\copyright~CERN on behalf of the \lhcb collaboration, license \href{http://creativecommons.org/licenses/by/3.0/}{CC-BY-3.0}.}}
\vspace*{2mm}

\end{titlepage}


\newpage
\setcounter{page}{2}
\mbox{~}
\newpage

\input{LHCb_authorlist.tex}

\cleardoublepage

%% file: LHCb_authorlist.tex
\centerline{\large\bf LHCb collaboration}
\begin{flushleft}
\small
R.~Aaij$^{40}$, 
B.~Adeva$^{36}$, 
M.~Adinolfi$^{45}$, 
A.~Affolder$^{51}$, 
Z.~Ajaltouni$^{5}$, 
J.~Albrecht$^{9}$, 
F.~Alessio$^{37}$, 
M.~Alexander$^{50}$, 
S.~Ali$^{40}$, 
G.~Alkhazov$^{29}$
% ...

{\footnotesize \it
% ...
$ ^{5}$Clermont Universit\'{e}, Universit\'{e} Blaise Pascal, CNRS/IN2P3, LPC, Clermont-Ferrand, France\\
% ...
$ ^{9}$Fakult\"{a}t Physik, Technische Universit\"{a}t Dortmund, Dortmund, Germany\\
% ...
$ ^{29}$Petersburg Nuclear Physics Institute (PNPI), Gatchina, Russia\\
% ...
$ ^{36}$Universidad de Santiago de Compostela, Santiago de Compostela, Spain\\
$ ^{37}$European Organization for Nuclear Research (CERN), Geneva, Switzerland\\
% ...
$ ^{40}$Nikhef National Institute for Subatomic Physics, Amsterdam, The Netherlands\\
% ...
$ ^{45}$H.H. Wills Physics Laboratory, University of Bristol, Bristol, United Kingdom\\
% ...
$ ^{50}$School of Physics and Astronomy, University of Glasgow, Glasgow, United Kingdom\\
$ ^{51}$Oliver Lodge Laboratory, University of Liverpool, Liverpool, United Kingdom\\
% ...
}
\end{flushleft}

%% file: paper-introduction.tex
\def\parenbar#1{{\null\!                        
   \mathop#1\limits^{\hbox{\tiny (---)}}       	
   \!\null}}  
   
\section{Introduction}
\label{sec:Introduction}

Within the framework of heavy quark expansion 
(HQE) theory~\cite{Khoze:1983yp, Shifman:1984wx, Shifman:1986mx,
Bigi:1992su, Bigi:1995jr, Uraltsev:1998bk, Neubert:1997gu}, 
$b$-hadron observables are calculated as a perturbative expansion
in inverse powers of the 
$b$-quark mass, $m_b$. At zeroth order the lifetimes of all 
weakly decaying $b$ hadrons are equal, with
corrections appearing at order $1/m_b^2$.
Ratios of $b$-hadron lifetimes can be theoretically predicted
with higher accuracy than absolute lifetimes
since many terms in the HQE cancel.
The latest theoretical predictions and world-average values
for the $b$-hadron lifetimes and lifetime ratios are reported in
Table~\ref{tab:current_ratios}.
A measurement of the ratio of the $\Lb$ baryon lifetime,
using the $\Lb\to\jpsi pK^-$ decay
mode\footnote{Charge conjugation is implied throughout this paper, unless otherwise stated.},
to that of the \Bd meson lifetime has recently been made
by the \lhcb collaboration~\cite{LHCb-PAPER-2013-032} and is not yet 
included in the world average.

In this paper, a measurement of the lifetimes of the $\Bu$, $\Bd$ and $\Bs$ mesons
and $\Lb$ baryon is reported using $pp$ collision data, corresponding to an
integrated luminosity of 1.0\,fb$^{-1}$, collected in 2011 with the LHCb detector
at a centre-of-mass energy of $7$\tev. 
The lifetimes are measured from the 
reconstructed $b$-hadron decay time distributions of the exclusive decay modes
\BuToJPsiK, \BdToJPsiKstENT, \BdToJPsiKS, \BsToJPsiPhi and \LbToJPsiL.
Collectively, these are referred to as \bToJPsiX decays.
In addition, measurements of lifetime ratios are reported.

As a result of neutral meson mixing the decay time distribution of
neutral $B^0_{q}$ mesons ($q\in\{s,d\}$) is characterised by two parameters, namely the
average decay width $\Gamma_{q}$ and the decay width difference
$\Delta\Gamma_{q}$ between the light (L) and heavy (H) $B^0_{q}$ mass eigenstates.
The summed decay rate of $B^0_{q}$ and $\overline{B}^0_{q}$ mesons
to a final state $f$ is given by~\cite{Hartkorn:1999ga, Dunietz:2000cr, Fleischer:2011cw}
\begin{equation}
\langle\Gamma(B^0_{q}(t)\to f)\rangle \equiv
\Gamma(B^0_{q}(t)\to f) + \Gamma(\overline{B}^0_{q}(t)\to f)
= R^f_{q,\rm L}e^{-\Gamma_{q,\rm L}t} + R^f_{q,\rm H}e^{-\Gamma_{q,\rm H}t},
\end{equation}
where terms proportional to the small flavour specific asymmetry, $a_{\rm fs}^q$, are
ignored~\cite{Nierste:2004uz}.
Therefore, for non-zero $\Delta\Gamma_{q}$ the decay time distribution of neutral
$B^0_{q}$ decays is not purely exponential. In the case of an equal admixture of 
$B^0_{q}$ and $\overline{B}^0_{q}$ at $t=0$, the observed average decay time is
given by~\cite{Fleischer:2011cw}
\begin{equation}
\tau_{B_q^0 \rightarrow f} = \frac{1}{\Gamma_q}\frac{1}{1-y_q^2} \left( \frac{1+2\mathcal{A}^f_{\Delta \Gamma_q} y_q + y_q^2}{1+\mathcal{A}^f_{\Delta \Gamma_q} y_q} \right),
\label{eqn:single}
\end{equation}
where $y_q \equiv \Delta\Gamma_q/(2\Gamma_q)$ 
and
$\mathcal{A}^f_{\Delta \Gamma_q}\equiv (R_{q,\rm H}^f - R_{q,\rm L}^f)/(R_{q,\rm H}^f + R_{q,\rm L}^f)$
is an observable that depends on the final state, $f$.
As such, the lifetimes measured are usually referred to as {\it effective} lifetimes.
In the \Bs system, where 
$\Delta\Gamma_s/\Gamma_s = 0.159 \pm 0.023$~\cite{HFAG}, the
deviation from an exponential decay time distribution is non-negligible.
In contrast, in the \Bd system this effect is expected to be small as 
$\Delta\Gamma_d/\Gamma_d$ is
predicted to be $(42\pm 8)\times10^{-4}$ in the Standard Model (SM)~\cite{Lenz:2006hd,Lenz:2011ti}.
Both the BaBar~\cite{Aubert:2003hd, Aubert:2004xga} and
Belle~\cite{Higuchi:2012kx} collaborations have measured
$|\Delta\Gamma_d/\Gamma_d|$ and the current world average is
$|\Delta\Gamma_d/\Gamma_d| = 0.015 \pm 0.018$~\cite{HFAG}. 
A deviation in the value of $\Delta\Gamma_d$ from the SM
prediction has recently been
proposed~\cite{Borissov:2013wwa} as a potential explanation for the 
anomalous like-sign dimuon charge asymmetry measured by
the D0 collaboration~\cite{Abazov:2013uma}.
In this paper, $\Delta\Gamma_d/\Gamma_d$ is measured
from the effective lifetimes of \BdToJPsiKstENT and \BdToJPsiKS decays, as
proposed in Ref.~\cite{Gershon:2010wx}.

The main challenge in the measurements
reported is understanding and controlling the detector acceptance,
reconstruction and selection efficiencies that depend upon the
$b$-hadron decay time. This paper is organised as follows.
Section~\ref{sec:Detector} describes the \lhcb detector and software. 
The selection criteria for the $b$-hadron
candidates are described in Sec.~\ref{sec:selection}.
Section~\ref{sec:efficiency} describes the reconstruction efficiencies
and the techniques used to correct the decay time distributions.
Section~\ref{sec:fit} describes how the efficiency corrections are incorporated into
the maximum likelihood fit that is used to measure the signal yields and
lifetimes. The systematic uncertainties on the measurements are 
described in Sec.~\ref{sec:systematics}. The final results and conclusions are
presented in Sec.~\ref{sec:results}. 

\begin{table}[t]
\caption{\small Theoretical predictions
and current world-average values~\cite{HFAG} for
$b$-hadron lifetimes and lifetime ratios.}
\label{tab:current_ratios}
\centerline{
\begin{tabular}{clc}
Observable		& 	\hfill Prediction			&	World average	\\
\hline
$\tau_{\Bu}$[\ps]	&	\hfill	--			&		$1.641 \pm 0.008$\\
$\tau_{\Bd}$[\ps]	&	\hfill	--			&		$1.519 \pm 0.007$\\
$\tau_{\Bs}$[\ps]	&	\hfill	--			&		$1.516 \pm 0.011$\\
$\tau_{\Lb}$[\ps]	&	\hfill	--			&		$1.429 \pm 0.024$\\
$\tau_{\Bu}/\tau_{\Bd}$ &	$1.063 		    \pm 0.027$~\cite{Beneke:2002rj, Franco:2002fc, Lenz:2011ti}
							 		&	$1.079 \pm 0.007$\\
$\tau_{\Bs}/\tau_{\Bd}$ &	$1.00\phantom{0} \pm 0.01\phantom{0}$~\cite{Beneke:1996gn, Keum:1998fd, Franco:2002fc, Lenz:2011ti}
									&	$0.998 \pm 0.009$\\
$\tau_{\Lb}/\tau_{\Bd}$ &	0.86--0.95\phantom{0}\ ~\cite{Shifman:1986mx, Uraltsev:1996ta, Bigi:1997fj, Pirjol:1998ur, Voloshin:1999ax, Tarantino:2003qw, Franco:2002fc, Gabbiani:2003pq, Gabbiani:2004tp} 	
									& 	$0.941 \pm 0.016$\\
\hline
\end{tabular}
}
\end{table}

%% file: paper-detector.tex
\section{Detector and software}
\label{sec:Detector}

The \lhcb detector~\cite{Alves:2008zz} is a single-arm forward
spectrometer covering the \mbox{pseudorapidity} range $2<\eta <5$,
designed for the study of particles containing \bquark or \cquark
quarks. The detector includes a high-precision tracking system
consisting of a silicon-strip vertex detector (VELO) surrounding the $pp$
interaction region, a large-area silicon-strip detector (TT)
located upstream of a dipole magnet with a bending power of about
$4{\rm\,Tm}$, and three stations of silicon-strip detectors and straw
drift tubes placed downstream.
The combined tracking system provides a momentum, $p$, measurement with
relative uncertainty that varies from 0.4\% at 5\gevc to 0.6\% at 100\gevc,
and impact parameter resolution of 20\mum for
charged particles with high transverse momentum, $p_{\rm T}$. 
Charged hadrons are identified
using two ring-imaging Cherenkov detectors~\cite{LHCb-DP-2012-003}. Photon, electron and
hadron candidates are identified by a calorimeter system consisting of
scintillating-pad and preshower detectors, an electromagnetic
calorimeter and a hadronic calorimeter. Muons are identified by a
system composed of alternating layers of iron and multiwire
proportional chambers~\cite{LHCb-DP-2012-002}.
The right-handed coordinate system adopted has the 
$z$-axis along the beam line and the $y$-axis along the vertical.
The trigger~\cite{LHCb-DP-2012-004} consists of a
hardware stage, based on information from the calorimeter and muon
systems, followed by a software stage, which applies a full event
reconstruction.

Two distinct classes of tracks are reconstructed using
hits in the tracking stations on both sides of the magnet, either with hits in the
VELO (long track) or without (downstream track). The vertex resolution of
$b$-hadron candidates reconstructed using long tracks is better than that for 
candidates reconstructed using downstream tracks. However, the use of
long tracks introduces a
dependence of the reconstruction efficiency on the $b$-hadron decay time.

In the simulation, $pp$ collisions are generated using
\pythia~6.4~\cite{Sjostrand:2006za} with a specific \lhcb
configuration~\cite{LHCb-PROC-2010-056}.  Decays of hadronic particles
are described by \evtgen~\cite{Lange:2001uf}, in which final state
radiation is generated using \photos~\cite{Golonka:2005pn}. The
interaction of the generated particles with the detector and its
response are implemented using the \geant
toolkit~\cite{Allison:2006ve, *Agostinelli:2002hh} as described in
Ref.~\cite{LHCb-PROC-2011-006}.

%% file: paper-selection.tex
\section{Candidate selection}
\label{sec:selection}

The reconstruction of each of the \bToJPsiX decays is similar and 
commences by selecting \mbox{$\jpsi\to\mu^{+}\mu^{-}$} decays. 
Events passing the hardware trigger contain dimuon
candidates with high transverse momentum.
The subsequent software trigger is composed of two stages.
The first stage performs a partial event reconstruction and requires events to 
have two well-identified oppositely charged muons with an invariant mass larger than $2.7\gevcc$. 
The selection at this stage has a uniform efficiency as a function of decay time.
The second stage performs a full event reconstruction, calculating
the position of each $pp$ interaction vertex (PV) using all available charged
particles in the event. The average number of PVs in each event is approximately $2.0$.
Their longitudinal ($z$) position is known to a
precision of approximately $0.05\mm$.
If multiple PVs are reconstructed in the event, the one with the
minimum value of $\chi^2_{\rm IP}$ is associated with the \jpsi candidate,
where $\chi^2_{\rm IP}$ is the increase in the $\chi^2$ of the PV fit if
the candidate trajectory is included.
Events are retained for further processing if they contain a $\jpsi\to\mu^{+}\mu^{-}$ pair
that forms a vertex that is significantly displaced from the PV.
This introduces a non-uniform efficiency as function of 
decay time.
 
The offline sample of \jpsi meson candidates is selected by
requiring each muon to have $p_{\rm T}$ larger than 500\mevc and the \jpsi 
candidate to be displaced from the PV by more than three times its decay length uncertainty.
The invariant mass of the two muons, $m(\mu^+\mu^-)$, must be in the range $[3030, 3150]\mevcc$. 

The $b$-hadron candidate selection is performed by applying kinematic and
particle identification criteria to the final-state tracks, the details of which are reported in
Sec.~\ref{sec:Bu} to~\ref{sec:Lb}.
No requirements are placed on variables that are highly correlated to the $b$-hadron
decay time, thereby avoiding the introduction of additional biases.
All final-state particles are required to have a pseudorapidity in the range
$2.0 < \eta < 4.5$. In addition, the $z$-position of the PV ($z_{\rm PV}$) is required to be
within $100\mm$ of the nominal interaction point, where the standard deviation of the 
$z_{\rm PV}$ distribution is approximately $47\mm$.
These criteria cause a reduction of approximately $10\%$ in signal yield but define
a fiducial region where the reconstruction efficiency is
largely uniform.

The maximum likelihood fit uses the invariant mass, $m(\jpsi X)$, and proper decay time, $t$,
of each $b$-hadron candidate. 
The decay time of the $b$-hadron candidate in its rest frame is derived from the relation $t = m\, l/q$,
where $m$ is its invariant mass and the decay length, $l$,
and the momentum, $q$, are measured in the experimental frame.
In this paper, $t$ is computed using a kinematic decay-tree fit (DTF)~\cite{Hulsbergen:2005pu}
involving all final-state tracks from the $b$-hadron candidate with a constraint on the position 
of the associated PV. 
Unlike in the trigger, the position of each PV is calculated using all
available charged particles in the event after the removal of the
$b$-hadron candidate final-state tracks. This is necessary to prevent 
the final-state tracks from biasing the PV position towards the $b$-hadron decay vertex
and helps to reduce the tails of the decay-time resolution function.  
This prescription does not bias the measured lifetime using simulated events.
The $\chi^{2}$ of the fit, $\chi^2_{\rm DTF}$, is useful
to discriminate between signal and background. In cases where there are multiple $b$-hadron
candidates per event, the candidate with the smallest $\chi^{2}_{\rm DTF}$ is chosen.
The $z$-position of the displaced $b$-hadron vertices are known to a
precision of approximately $0.15\mm$.

Studies of simulated events 
show that in the case of \BdToJPsiKst (\BsToJPsiPhi) decays,
imposing requirements on $\chi^2_{\rm DTF}$
introduces a dependence of the selection efficiency on the decay time
if the \Kp and $\pi^-$ (\Kp and \Km) tracks are included in the DTF.
If no correction is applied to the decay time distribution,
the measured lifetime would be biased
by approximately $-2\fs$ relative to the generated value.
Using simulated events it is found that this effect is correlated to the 
opening angle between the \Kp and $\pi^-$ (\Kp and \Km) from the $K^{*0}$ ($\phi$)
decay.
No effect is observed for the muons coming from the \jpsi decay due to the larger
opening angle in this case. 
To remove the effect, the calculation of $\chi^2_{\rm DTF}$ 
for the \BdToJPsiKst and \BsToJPsiPhi channels
is performed with an alternative DTF in which the assigned track
parameter uncertainties of the kaon and pion are increased in such a way that
their contribution to the $b$-hadron vertex position is negligible.

Candidates are required to have $t$ in the range $[0.3, 14.0]\ps$. The lower
bound on the decay time suppresses a large fraction of the prompt combinatorial background
that is composed of tracks from the same PV, while the upper bound is introduced to
reduce the sensitivity to long-lived background candidates.
In the case of the \BdToJPsiKS and \LbToJPsiL decays, the lower bound is increased to 0.45\ps
to compensate for the worse decay time resolution in these modes. 

In events with multiple PVs, $b$-hadron candidates are 
removed if they have a $\chi^2_{\rm IP}$ with respect to the next best
PV smaller than $50$. This requirement is found to 
distort the decay time distribution, but reduces 
a source of background due to the incorrect association of the $b$ hadron to its production PV.

\begin{figure}[h!]
\centering
\begin{overpic}[scale=0.75]{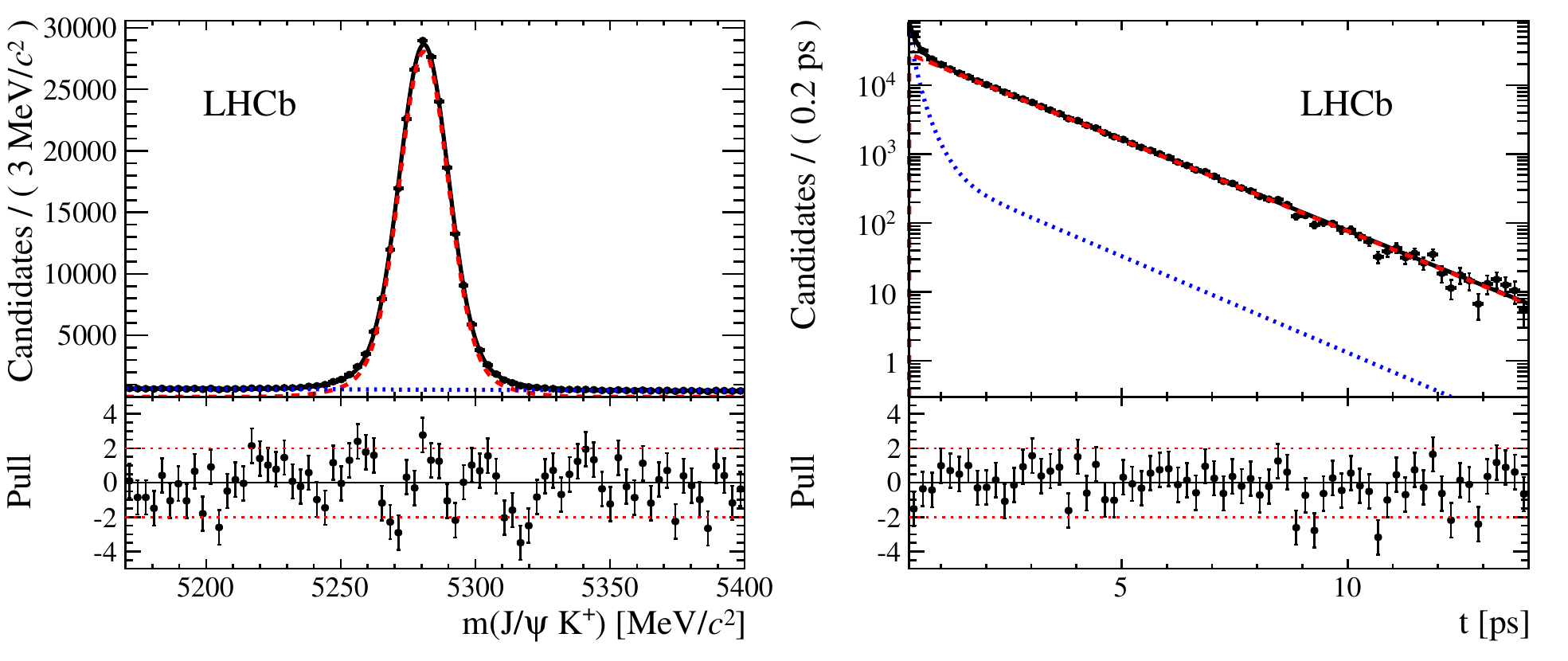}
\put(11,31){\scriptsize \BuToJPsiK}
\put(81,31){\scriptsize \BuToJPsiK}
\end{overpic}
\begin{overpic}[scale=0.75]{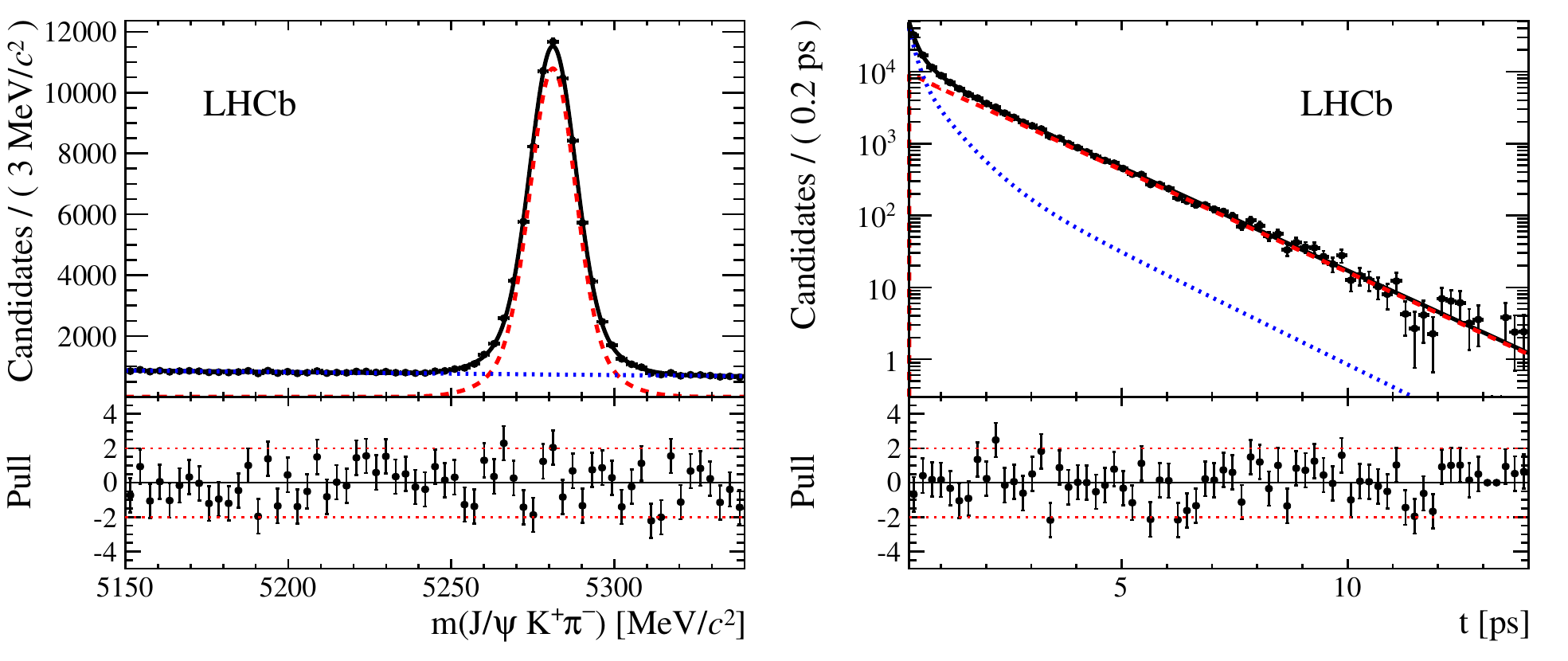}
\put(11,31){\scriptsize \BdToJPsiKst}
\put(81,31){\scriptsize \BdToJPsiKst}
\end{overpic}
\begin{overpic}[scale=0.75]{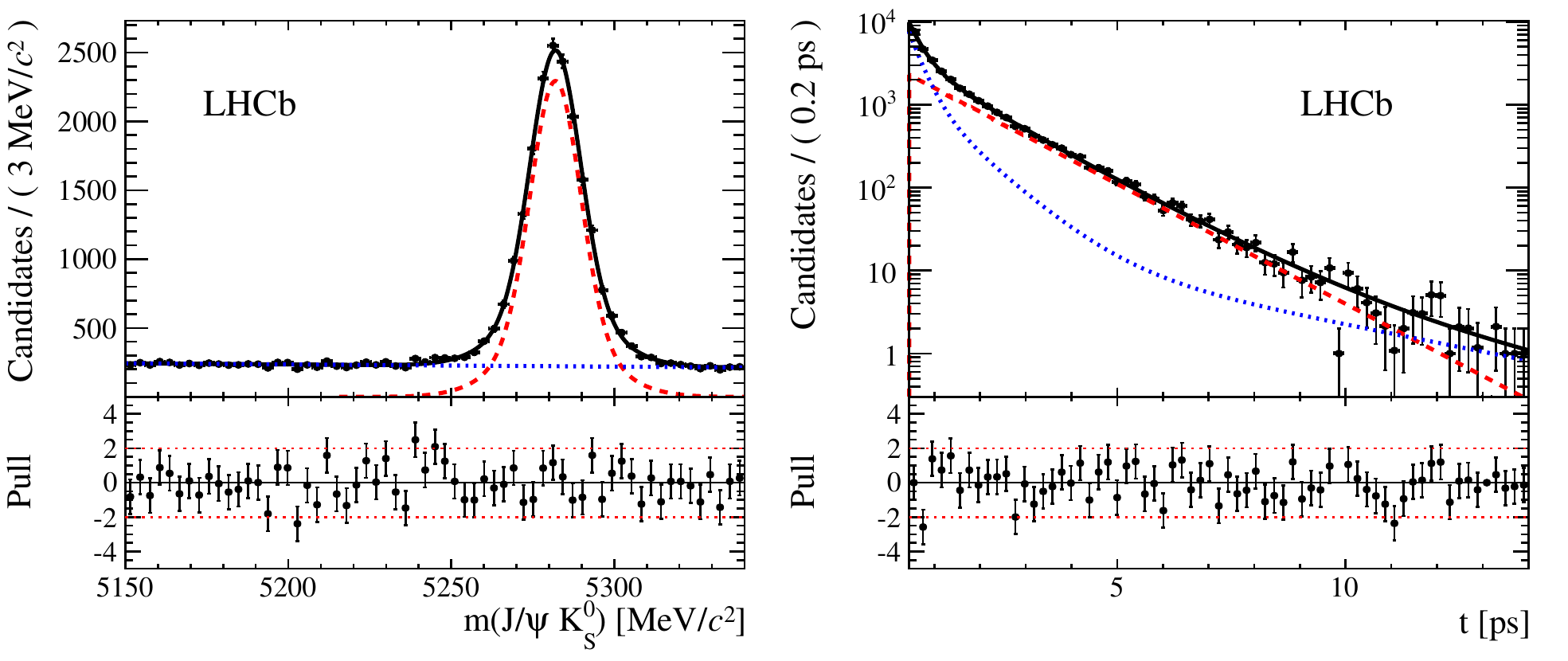}
\put(11,31){\scriptsize \BdToJPsiKS}
\put(81,31){\scriptsize \BdToJPsiKS}
\end{overpic}
\caption{Distributions of the (left) mass and (right) decay time of 
\BuToJPsiK, \BdToJPsiKst and \BdToJPsiKS candidates and their associated
residual uncertainties (pulls). The data are shown
by the black points; the total fit function by the black solid line; the signal contribution by the red dashed line and
the background contribution by the blue dotted line. 
\label{fig:B_cfit}}
\end{figure}

\begin{figure}[h]
\centering
\begin{overpic}[scale=0.75]{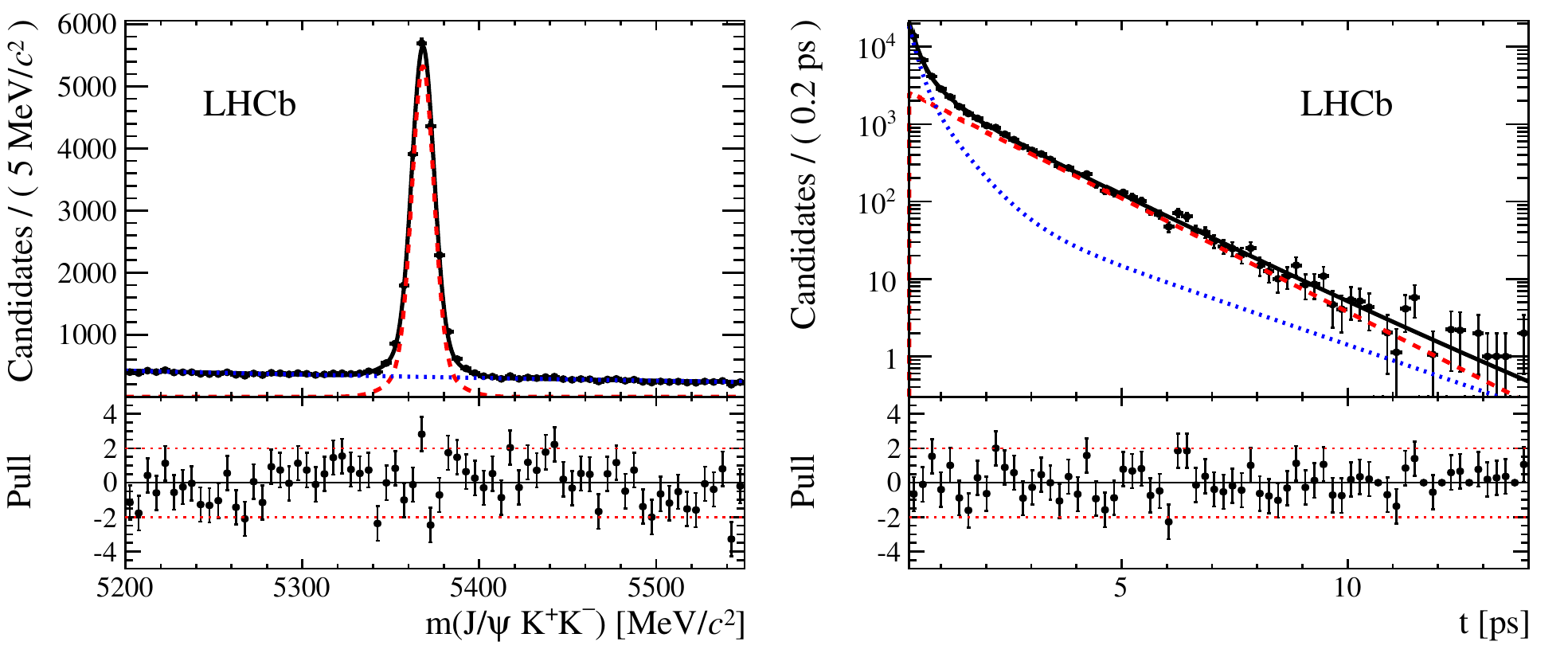}
\put(11,31){\scriptsize \BsToJPsiPhi}
\put(81,31){\scriptsize \BsToJPsiPhi}
\end{overpic}
\begin{overpic}[scale=0.75]{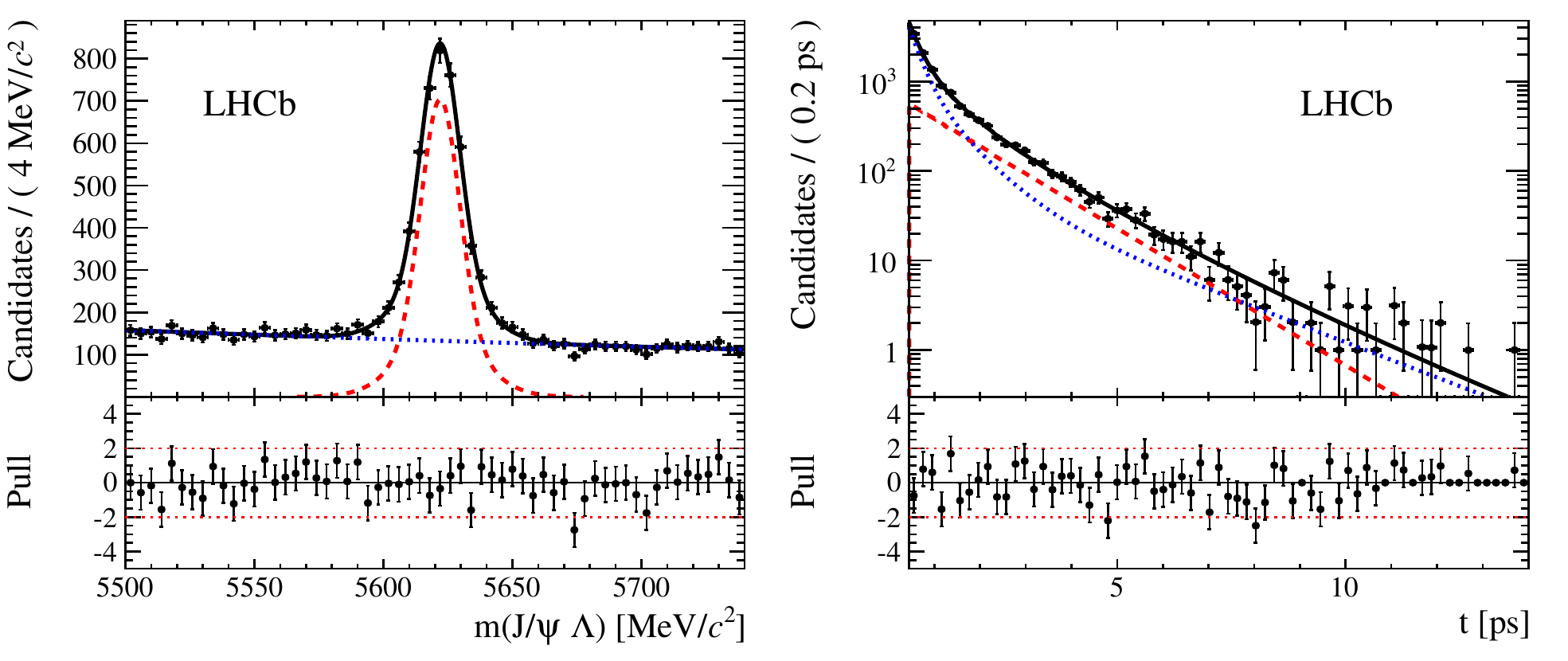}
\put(11,31){\scriptsize $\Lb\to\jpsi\Lz$}
\put(81,31){\scriptsize $\Lb\to\jpsi\Lz$}
\end{overpic}
\caption{Distributions of the (left) mass and (right) decay time of \BsToJPsiPhi and $\Lb\to\jpsi\Lz$
candidates and their associated
residual uncertainties (pulls). The data are shown
by the black points; the total fit function by the black solid line; the signal contribution by the red dashed line and
the background contribution by the blue dotted line.  \label{fig:Bs_cfit}}
\end{figure}

The invariant mass is computed using another
kinematic fit without any constraint on the PV position but with the invariant mass
of the $\mu^{+}\mu^{-}$ pair, $m(\mu^+\mu^-)$,
constrained to the known \jpsi\ mass~\cite{PDG2012}.
Figures~\ref{fig:B_cfit} and~\ref{fig:Bs_cfit} show the
$m(\jpsi X)$ distributions for the selected 
candidates in each final state and Table~\ref{tab:yields} gives the corresponding signal yields. 

\clearpage

\begin{table}[!]
\caption{\small Estimated event yields for the five $b \to \jpsi X$ channels selected using the criteria 
described in Sec.~\ref{sec:Bu} to~\ref{sec:Lb}.}  
\begin{center}
\begin{tabular}{lr}
\hspace{0.4cm}Channel						&	Yield	\hspace{0.6cm}	\\\hline
\BuToJPsiK						&	 $229\, 434\pm 503$ 		\\
\BdToJPsiKst						&	 $70\,534\pm 312$  		\\
\BdToJPsiKS						&	 $17\,045 \pm 175$  	 \\
\BsToJPsiPhi						&	 $18\,662\pm 152$  	 	\\
$\LbToJPsiL$						&	 $3\,960 \pm\ \, 89$  	 \\
\hline
\end{tabular}

\label{tab:yields}
\end{center}
\end{table}

\subsection{Selection of \boldmath{\BuToJPsiK} decays\label{sec:Bu}}

The \Bu candidates are reconstructed by combining the \jpsi candidates with a charged 
particle that is identified as a kaon with $\pt$ larger than $1\gevc$ 
and $p$ larger than $10\gevc$. 
The invariant mass, $m(\jpsi\Kp)$, must be in the range $[5170, 5400]\mevcc$,
where the lower bound is chosen to
remove feed-down from incompletely reconstructed $\BdToJPsiKst$ decays.
The $\chi^{2}_{\rm DTF}$ of the 
fit, which has 5 degrees of freedom, is required to be less than 25. 
Multiple \Bu candidates are found in less than $0.02\%$ of selected events. 


\subsection{Selection of \boldmath{\BdToJPsiKst} decays\label{sec:sel_Bd}}
 
The  $K^{*0}$  candidates are reconstructed by combining two oppositely charged particles
that are identified as a kaon and a pion.  The pion and $K^{*0}$ 
must have $\pt$ greater than $0.3\gevc$ and $1.5\gevc$, respectively.
The invariant mass, $m(\Kp\pi^-)$, must be in the range $[826, 966]\mevcc$.

The \Bd candidates are reconstructed by combining the \jpsi and  $K^{*0}$  candidates.
The invariant mass, $m(\jpsi\Kp\pi^-)$, must be in the range $[5150, 5340]\mevcc$,
where the upper bound is chosen to remove the contribution from 
$\Bs\to\jpsi \overline{K}^{*0}$ decays.
The $\chi^{2}_{\rm DTF}$ of the fit, which has 3 degrees of freedom,
is required to be less than 15. 
Multiple \Bd candidates are found in $2.2\%$ of selected events.

\subsection{Selection of \boldmath{\BdToJPsiKS} decays\label{sec:BdJpsiKS}}

The \KS candidates are formed from the combination of
two oppositely charged particles that are identified as pions and
reconstructed as downstream tracks. This 
is necessary since studies of simulated signal decays demonstrate that 
an inefficiency depending on
the $b$-hadron decay time is introduced by the reconstruction of the long-lived
\KS and $\Lz$ particles using long tracks. Even so, it is found that the
acceptance of the TT still depends on the origin of the tracks. This effect
is removed by further tightening of the requirement on the position of the
PV  to $z_{\rm PV} > -50\mm$.

For particles produced close to the interaction region, this effect is suppressed by the 
requirements on the fiducial region for the PV, which is further tightened
by requiring that , to account for the additional acceptance
introduced by the TT.

The downstream pions are required to have $\pt$ greater than 
$0.1\gevc$ and $p$ greater than $2\gevc$. The \KS 
candidate must have $\pt$ greater than $1\gevc$ and 
be well separated from the \Bd decay vertex, to suppress potential background
from \BdToJPsiKst decays where the kaon has been misidentified as a pion. The
$\chi^2$ of the \KS vertex fit must be less than 25 and the invariant mass of the dipion system,
$m(\pi^+\pi^-)$, must be within $15\mevcc$ of the known \KS mass~\cite{PDG2012}.
For subsequent stages of the selection, $m(\pi^+\pi^-)$ is constrained to the known \KS mass.

The invariant mass, $m(\jpsi\KS)$, of the \jpsi and \KS candidate combination must be in
the range $[5150, 5340]\mevcc$,
where the upper bound is chosen to remove the contribution from 
$\Bs\to\jpsi\KS$ decays. 
The $\chi^{2}_{\rm DTF}$ of the 
fit, which has 6 degrees of freedom, is required to be less than 30. 
Multiple \Bd
candidates are found in less than $0.4\%$ of selected events.



\subsection{Selection of \boldmath{\BsToJPsiPhi} decays\label{sec:Bs}}
The $\phi$ candidates are formed from two oppositely charged particles that have been identified as
kaons and originate from a common vertex. The $\Kp\Km$ pair  is required to have $p_{\rm T}$
larger than $1\gevc$.
The invariant mass of the $\Kp\Km$ pair, $m(\Kp\Km)$, must be in the 
range $[990, 1050]\mevcc$. 

The \Bs candidates are reconstructed by combining the \jpsi\ candidate with the $\Kp\Km$ pair, requiring the 
invariant mass, $m(\jpsi\Kp\Km)$, to be in the range $[5200, 5550]\mevcc$. 
The $\chi^{2}_{\rm DTF}$ of the 
fit, which has 3 degrees of freedom, is required to be less than 15. Multiple \Bs candidates are 
found in less than $2.0\%$ of selected events.

\subsection{Selection of \boldmath{\LbToJPsiL} decays\label{sec:Lb}}

The selection of \LbToJPsiL candidates follows a similar approach to that adopted
for \BdToJPsiKS decays.
Only downstream protons and pions are used to reconstruct the $\Lz$
candidates. The pions are required to have 
$\pt$ larger than $0.1\gevc$, while pions and protons must have $p$ larger than 
$2\gevc$. The $\Lz$ 
candidate must be well separated from the $\Lb$ decay vertex and have 
$\pt$ larger than $1\gevc$. The
$\chi^2$ of the $\Lz$ vertex fit must be less than 25 and $m(p\pi^-)$ 
must be within $6\mevcc$ of the known $\Lz$ mass~\cite{PDG2012}.
For subsequent stages of the selection, $m(p\pi^-)$ is constrained to the known
$\Lz$ mass.

The  invariant mass, $m(\jpsi\Lz)$, of the \jpsi and $\Lz$  candidate 
combination must be in the range $[5470, 5770]\mevcc$.
The $\chi^{2}_{\rm DTF}$ of the 
fit, which has 6 degrees of freedom, is required to be less than 30. 
Multiple $\Lb$
candidates are found in less than $0.5\%$ of selected events.

%% file: paper-efficiency.tex
\section{Dependence of efficiencies on decay time}
\label{sec:efficiency}

Section~\ref{sec:selection} described the reconstruction and selection criteria of 
the \bToJPsiX decays and various techniques that have been used to minimise
the dependence of selection efficiencies upon the decay time. After
these steps, there remain two effects that distort the $b$-hadron decay time
distribution. These are caused by the VELO-track reconstruction efficiency,
$\varepsilon_{\rm VELO}$, and the combination of the trigger efficiency, $\varepsilon_{\rm trigger}$,
and offline selection efficiency, $\varepsilon_{\rm selection|trigger}$. 
This section will describe these effects and
the techniques that are used to evaluate the efficiencies from data control samples.

\subsection{VELO-track reconstruction efficiency\label{sec:velo}}

The largest variation of the efficiency with the decay time
is introduced by the track reconstruction in the VELO. 
The track finding procedure in the VELO assumes that
tracks originate approximately from the interaction region~\cite{Alves:2008zz, LHCb-PUB-2011-001}.
In the case of long-lived $b$-hadron candidates this assumption
is not well justified, leading to a loss of reconstruction efficiency for charged
particle tracks from the $b$-hadron decay.

\begin{figure}[t]
\centering{
\begin{overpic}[scale=0.39]{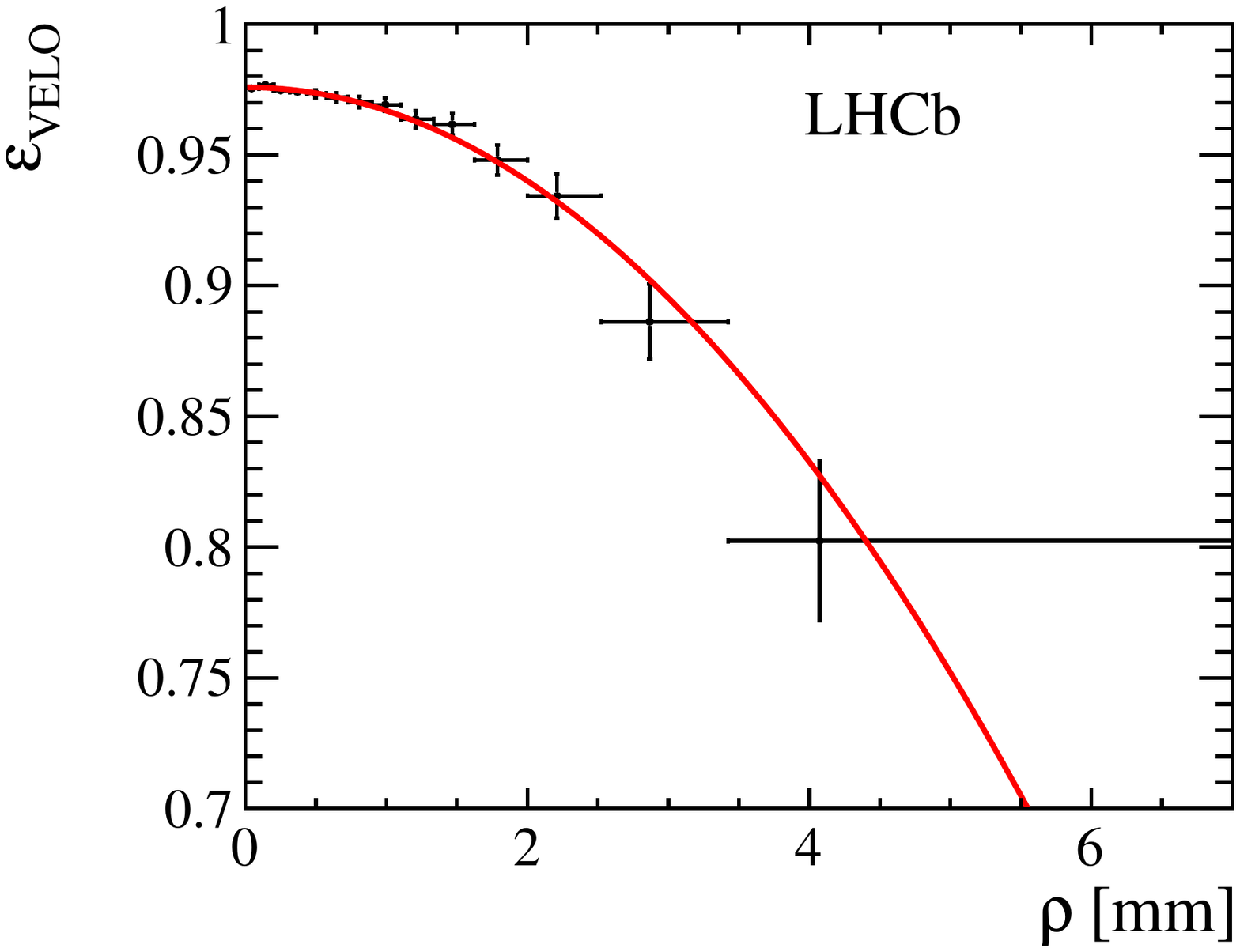}
\put(79,59.7){(a)}
\end{overpic}
\begin{overpic}[scale=0.39]{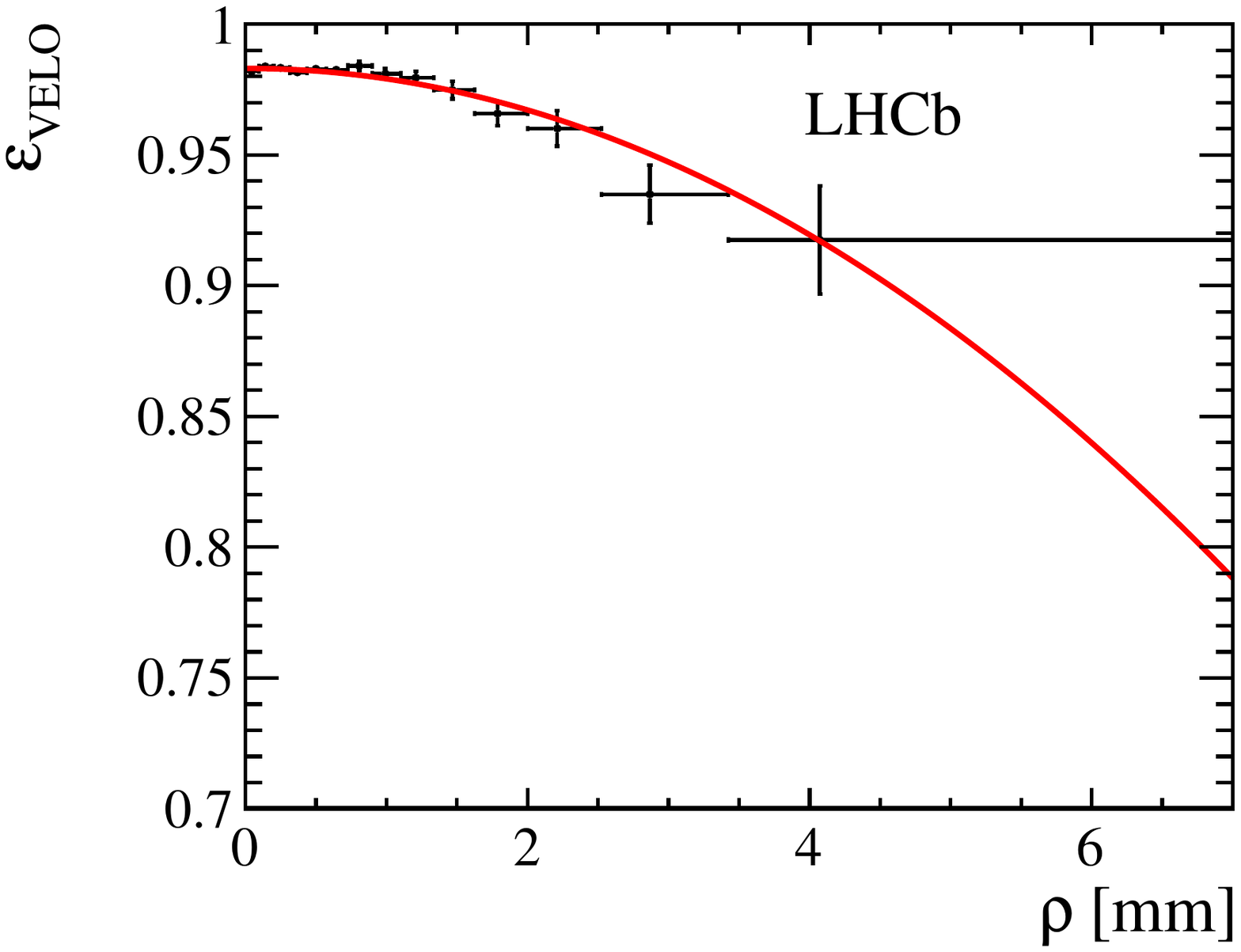}
\put(79,59.7){(b)}
\end{overpic}
}
\caption{\small VELO-track reconstruction efficiency for kaon tracks reconstructed using the (a) online and (b) offline algorithms as a function of the kaon \DOCAz, as defined in Eq.~(\ref{eqn:docaz}). The red solid lines show the result of an unbinned maximum likelihood fit using the parameterisation in Eq.~(\ref{eqn:docaz_param}) to the background subtracted data (black points).\label{fig:docaz_eff}}
\end{figure}

The distance of closest approach of the track to the $z$-axis is defined as
\begin{eqnarray}
\DOCAz &\equiv& \frac{\left| (\boldsymbol{d}-\boldsymbol{v})\cdot(\boldsymbol{p}\times\boldsymbol{\hat{z}})\right|}{ \left| \boldsymbol{p}\times{\boldsymbol{\hat{z}}}\right|},
\label{eqn:docaz}
\end{eqnarray}
where $\boldsymbol{p}$ is the momentum of the final-state track from a
$b$-hadron candidate decaying at point $\boldsymbol{d}$, $\boldsymbol{\hat{z}}$
is a unit vector along the $z$-axis and $\boldsymbol{v}$ is the origin
of the VELO coordinate system. During data taking the position of
the \lhcb VELO is monitored as a function of time and is centred around the LHC beam line.
Using a control sample of \mbox{\BuToJPsiK} candidates where the 
\Kp is reconstructed as a downstream track, the VELO-track reconstruction
efficiency, $\varepsilon_{\rm VELO}(\DOCAz)$,
is computed as the fraction of these tracks that are
also reconstructed as long tracks. From samples of simulated
$b$-hadron decays, it is observed that $\varepsilon_{\rm VELO}(\DOCAz)$ 
can be empirically parameterised by
\begin{equation}
\varepsilon_{\rm VELO}(\DOCAz) = a  ( 1 + c\DOCAz^{2}),
\label{eqn:docaz_param}
\end{equation}
where the parameters $a$ and $c$ are determined
from a fit to the unbinned efficiency distribution.

\begin{table}[t]
\caption{\small VELO reconstruction efficiency in data for kaon tracks reconstructed with the
online and offline algorithms. In both cases, the
correlation coefficient between $a$ and $c$ is 0.2.\label{tab:data_weights}}
\centering{
\begin{tabular}{ccc}
			&	$a$					& $c$ [$\mm^{-2}$] 		\\\hline
Online		& 		$0.9759 \pm 0.0005$ 	& $ -0.0093 \pm 0.0007$ 	 \\
Offline		& 		$0.9831 \pm 0.0004$ 	& $ -0.0041 \pm 0.0005$ 	 \\
\hline
\end{tabular}
}
\end{table}

Figure~\ref{fig:docaz_eff} shows the VELO-track reconstruction efficiency obtained using this method 
and Table~\ref{tab:data_weights} shows the corresponding fit results.
Since different configurations of the VELO reconstruction algorithms are used within
the \lhcb software trigger (online) and during the subsequent processing 
(offline), it is necessary to evaluate two different efficiencies.
The stronger dependence of the online efficiency as a function of \DOCAz is due to the
additional requirements used in the first stage of the software
trigger such that it satisfies the required processing time.

Applying the same technique to a simulated sample of \BuToJPsiK decays
yields qualitatively similar behaviour
for $\varepsilon_{\rm VELO}(\DOCAz)$.
Studies on simulated data show that the efficiency for kaons and pions
from the decay of $\phi$ and $K^{*0}$ mesons is smaller than for the
kaon in \BuToJPsiK decays, due to the small opening
between the particles in the $\phi$ and $K^{*0}$ decays, as discussed in 
Sec.~\ref{sec:selection}. In addition, there are  kinematic
differences between the calibration \Bu sample and the signal samples.
Scaling factors on the efficiency parameters are derived from simulation to account for
these effects, and have typical sizes in the range $[1.04, 1.65]$, 
depending on the decay mode and final-state particle being considered.

The distortion to the $b$-hadron candidate decay time distribution caused by
the VELO-track reconstruction is corrected for by weighting each $b$-hadron candidate
by the inverse of the product of the per-track efficiencies. 
The systematic effect introduced by this weighting is tested using simulated samples of each channel.
The chosen efficiency depends on whether the particle is reconstructed with the online or offline
variant of the algorithm. Studies on simulated data show
that tracks found by the online tracking algorithm are also
found by the offline tracking efficiency.
For example, the efficiency weight for each \BdToJPsiKst candidate takes the form
\begin{eqnarray}
w_{\BdToJPsiKst} &=& 1/\left(\varepsilon^{\mu^{+}}_{\rm VELO, online}\ \varepsilon^{\mu^{-}}_{\rm VELO, online}\ \varepsilon^{K^{+}}_{\rm VELO, offline} \ \varepsilon^{\pi^{-}}_{\rm VELO, offline} \right),
\label{eqn:weights_example}
\end{eqnarray}
since the two muons are required to be reconstructed online, 
while the kaons and the pions are reconstructed offline.

In the case of the 
\BdToJPsiKS and \LbToJPsiL channels, since no 
VELO information is used when reconstructing the \KS and $\Lz$ particles, the 
candidate weighting functions take the form
$w = 1/\left(\varepsilon^{\mu^{+}}_{\rm VELO, online}\ \varepsilon^{\mu^{-}}_{\rm VELO, online}\right)$.

\subsection{Trigger and selection efficiency\label{sec:trigger}}

The efficiency of the second
stage of the software trigger depends on the $b$-hadron decay time as it
requires that the $\jpsi$ meson is significantly displaced from the PV.
A parameterisation of this efficiency, $\varepsilon_{\rm trigger}(t)$, is obtained for each $b\to\jpsi X$ decay
mode by exploiting a corresponding sample of  $b\to\jpsi X$ candidates that are selected without
any displacement requirement. For each channel, the control sample corresponds
to approximately $40\%$ of the total number of signal
candidates. A maximum likelihood fit to the unbinned invariant mass distribution $m(\jpsi X)$
is performed to determine the 
fraction of signal decays that survive the decay-time biasing trigger
requirements as a function of decay time.

The same technique is used to determine the decay time efficiency 
of the triggered candidates caused by the offline selection,
$\varepsilon_{\rm selection|trigger}(t)$, which is introduced by 
the requirement on the detachment of the \jpsi mesons in the sample
used to reconstruct the $b$-hadron decays. 
The combined selection efficiency, $\varepsilon_{\rm selection}(t)$, is given by the product
of $\varepsilon_{\rm trigger}(t)$ and $\varepsilon_{\rm selection|trigger}(t)$.

\begin{figure}[t]
\centering{
\includegraphics[scale=0.39]{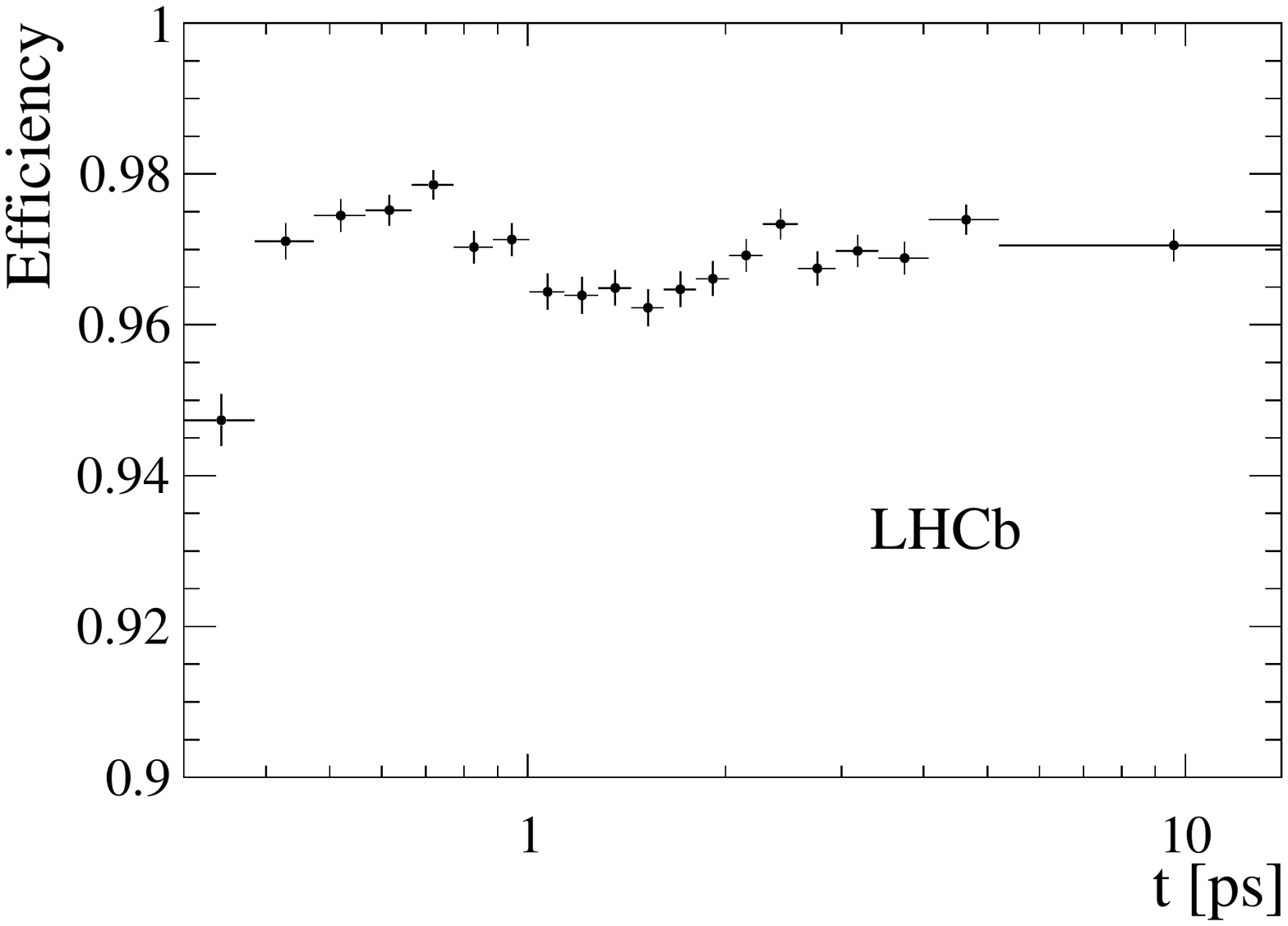}
}
\caption{\small Combined trigger and selection efficiency, $\varepsilon_{\rm selection}(t)$,
for \BuToJPsiK candidates.\label{fig:trig_eff}}
\end{figure}

Figure~\ref{fig:trig_eff} shows
$\varepsilon_{\rm selection}(t)$ obtained for the \BuToJPsiK
channel as a function of decay time.
The efficiencies obtained for the other \bToJPsiX channels are 
qualitatively similar. Studies using simulated events show that the
efficiency drop below $0.5\ps$ is caused by the $\jpsi$ displacement requirement.
The dip near $1.5\ps$ appears because the PV reconstruction in the
software trigger is such that
some final-state tracks of short-lived $b$-hadron decays may be used to reconstruct
an additional fake PV close to the true $b$-hadron decay vertex. As a result
the reconstructed \jpsi meson does not satisfy the displacement requirement, 
leading to a decrease in efficiency.

The efficiency parameterisation for each channel is used in the fit to measure the
corresponding $b$-hadron lifetime. An exception is made for the
$\Lb\to\jpsi\Lz$ channel where,
owing to its smaller event yield, $\varepsilon_{\rm selection}(t)$ measured with
$\BdToJPsiKS$ decays is used instead. The 
validity of this approach is checked using large samples of simulated events.

%% file: paper-fit.tex
\def\sigpdf{{\cal S}}
\def\bkgpdf{{\cal B}}
\def\fullpdf{{\cal P}}
\def\fsig{f_{s}}
\section{Maximum likelihood fit\label{sec:fit}}

For each channel, the lifetime is determined from a two-dimensional maximum likelihood fit to 
the unbinned $m(\jpsi X)$ and $t$ distributions.
The full probability density function (PDF) is constructed as
\mbox{$\fullpdf= \fsig ({\cal S}_m\times{\cal S}_t) + ( 1 - \fsig ) ({\cal B}_m\times{\cal B}_t)$},
where $\fsig$ is the signal fraction, determined in the fit,
and ${\cal S}_m\times{\cal S}_t$ and ${\cal B}_m\times{\cal B}_t$ are the
($m(\jpsi X)$, $t$) PDFs 
for the signal and background components, respectively.
A systematic uncertainty is assigned to the assumption that the PDFs
factorise.

The signal mass PDF,
$\sigpdf_m$,  is modelled by the sum of two Gaussian functions.
The free parameters in the fit are the common mean,
the width of the narrower Gaussian function,
the ratio of the second to the first Gaussian width
and the fraction of the first Gaussian function. 
The background mass distribution, $\bkgpdf_m$, is modelled by an
exponential function with a single free parameter.

The signal $b$-hadron decay time distribution is described by an exponential
function with decay constant given by the $b$-hadron lifetime, $\tau_{H_b\to\jpsi X}$.
The signal decay time PDF, ${\cal S}_t$, is obtained by multiplying the exponential function
by the combined $t$-dependent trigger and selection efficiency described in Sec.~\ref{sec:trigger}. 
From inspection of events in the sidebands of the $b$-hadron signal peak, the background 
decay time PDF, ${\cal B}_t$, is
well modelled by a sum of three exponential functions with different decay constants
that are free in the fit. 
These components originate from a combination of prompt 
candidates, where all tracks originate from the same PV, and long-lived candidates
where tracks from the associated PV are combined with other tracks of long-lived
particles. 
For each channel the exponential functions are convolved
with a Gaussian resolution function 
with width $\sigma$ and mean $\Delta$, an offset of the order of a few femtoseconds
that is fixed in the fit. 
Using a sample of prompt \jpsi background events, the decay time resolution 
for \bToJPsiX channels reconstructed using long tracks 
has been measured to be approximately $45\fs$~\cite{LHCb-PAPER-2013-002}.
For \BdToJPsiKS and \LbToJPsiL decays, which use downstream tracks to reconstruct
the \KS and $\Lz$ particles, a similar study of an event sample composed of
prompt \jpsi mesons combined
with two downstream tracks, reconstructed as either a \KS or \Lz, has determined the
resolution to be $65\fs$.
The systematic uncertainties related to the choice of resolution model
are discussed in Sec.~\ref{sec:systematics}.

The negative log-likelihood, constructed as
\begin{equation}
-\ln{\cal L} =  -  \alpha \sum_{\mathrm{events}\; i}{ w_i \ln{\fullpdf} },
\label{eqn:LL}
\end{equation}
is minimised in the fit, where the weights $w_i$
correspond to the per-candidate correction for the VELO reconstruction efficiency described in
Sec.~\ref{sec:velo}. The factor $\alpha = \sum_i w_i / \sum_i w_i^2$
is used to include the effect
of the weights in the determination of the uncertainties~\cite{2009arXiv0905.0724X}.
Figures~\ref{fig:B_cfit} and~\ref{fig:Bs_cfit} show the 
result of fitting this model to the selected candidates for each channel, projected onto the corresponding
$m(\jpsi X)$ and $t$ distributions.

As a consistency check, an alternative fit procedure is developed where each event is given a signal weight, 
$W_i$, determined using the \sPlot~\cite{splot} method with $m(\jpsi X)$ as the discriminating variable
and using the mass model described above. A weighted fit to the decay time distribution using the signal
PDF is then used to measure the $b$-hadron lifetime. In this case, the negative log-likelihood is given by
Eq.~(\ref{eqn:LL}) where $w_i$ is replaced with $W_i w_i$ and
$\alpha = \sum_i (W_i w_i) /  \sum_i (W_i w_i)^2$.
The difference
between the results of the two fitting procedures is used to estimate the
systematic uncertainty on the background description.

%% file: paper-systematics.tex
\section{Systematic uncertainties}
\label{sec:systematics}

The systematic effects affecting the measurements reported here are discussed in the following 
and summarised in Tables~\ref{tab:systematics} and~\ref{tab:systematics_ratios}.

The systematic uncertainty related to the VELO-track reconstruction efficiency can be split into
two components. The first
uncertainty is due to the finite size of the \BuToJPsiK sample, reconstructed 
using downstream kaon tracks, which is used to determine
the per-candidate efficiency weights and leads to a statistical uncertainty on the
$\varepsilon_{\rm VELO}(\DOCAz)$
parameterisation. The lifetime fits are
repeated after varying the parameters
by $\pm1\sigma$ and the largest difference between the lifetimes is 
assigned as the uncertainty.
The second uncertainty is due to the scaling factors, which are used to correct
the efficiency for phase-space effects,
obtained from simulated events. The fit is repeated using the unscaled efficiency
and half of the variation in fit results is assigned as a systematic uncertainty.
These contributions, of roughly the same size, are added in quadrature in Table~\ref{tab:systematics}. 

A number of additional consistency checks are performed to investigate
possible mismodelling of the VELO-track reconstruction efficiency. First, 
$\varepsilon_{\rm VELO}(\DOCAz)$ is evaluated in two track momentum and
two track multiplicity bins and the 
event weights recalculated. Using both data and simulated events, no significant change in the lifetimes 
is observed after repeating the fit with the updated weights and, therefore, no systematic uncertainty is assigned. 
Secondly, to assess the sensitivity to the choice of parameterisation for $\varepsilon_{\rm VELO}(\DOCAz)$
(Eq.~\ref{eqn:docaz_param}), the results are compared to those with linear model for the
efficiency. The effect is found to be negligible and no systematic uncertainty is applied. 
Thirdly, the dependence of the VELO-track reconstruction efficiency on the azimuthal angle, $\phi$, of each track 
is studied by independently evaluating the efficiency in four $\phi$ quadrants for both data and simulation. 
No dependence is observed.
Finally, the efficiency is determined separately for both positive and negative kaons
and found to be compatible.

The techniques described in Sec.~\ref{sec:efficiency} to correct the efficiency as a function of
the decay time are validated on simulated data. 
The lifetime is fit in each simulated signal mode and the departure from the generated lifetime,
$\Delta\tau$, is found to be compatible with zero within the statistical precision of each simulated sample.
The measured lifetimes in the data sample are corrected by each $\Delta\tau$ and a corresponding
systematic uncertainty is assigned, given by the size of the statistical uncertainty on the fitted 
lifetime for each simulated signal mode.

The assumption that $m(\jpsi X)$ is independent of the decay time is central to
the validity of the likelihood fits used in this study. It is tested by re-evaluating the signal
weights of the alternative fit in bins of decay time and then refitting the entire sample using the modified weights.
The systematic uncertainty for each decay mode is evaluated as 
the sum in quadrature of the lifetime variations, each weighted by the fraction of signal
events in the corresponding bin.

For each signal decay mode, the effect of the limited size of the
control sample used to estimate the combined trigger and selection efficiency
is evaluated by repeating the fits with $\varepsilon_{\rm selection}(t)$ randomly fluctuated
within its statistical uncertainty. The standard deviation of the distribution of lifetimes
obtained is assigned as the systematic uncertainty.

The alternative likelihood fit does not assume
any model for the decay time distribution associated with the 
combinatorial background. Therefore, the 
systematic uncertainty associated to the modelling of this background is evaluated by taking the difference
in lifetimes measured by the nominal and alternative fit methods.

The fit uses a double Gaussian function to describe the $m(\jpsi X)$ distribution of signal candidates. This 
assumption is tested by repeating the fit using a double-sided
Apollonios function~\cite{Santos:2013gra} where the mean and width parameters 
are varied in the fit and the remaining parameters are fixed to those from simulation.
The differences in
lifetime with respect to the default results are taken as systematic uncertainties.

As described in Sec.~\ref{sec:fit} the dominant background in each channel is combinatorial in nature. It
is also possible for backgrounds to arise due to misreconstruction of $b$-hadron decays where the particle
identification has failed. The presence of such backgrounds is checked by inspecting events in the
sidebands of the signal and re-assigning the mass hypotheses of at least one of the final-state hadrons. The
only contributions that have an impact are $\Lb\to\jpsi p\Km$ decays in the $\Bs\to\jpsi\phi$ channel where a
proton is misidentified as a kaon and the cross-feed component between  $\Bd\to\jpsi\KS$ and 
$\Lb\to\jpsi\Lz$ decays where pion and protons are misidentified.
In the case of $\Bs\to\jpsi\phi$ decays, the fit is repeated including
a contribution of $\Lb\to\jpsi p\Km$ decays. The two-dimensional PDF is determined
from simulation, while the yield is found to be $6\%$ from the sidebands of the \Bs sample.
This leads to the effective lifetime changing by $0.4\fs$, which 
is assigned as a systematic uncertainty. A similar procedure is used 
to determine the invariant mass shape of the cross-feed background between $\Bd\to\jpsi\KS$ 
and $\Lb\to\jpsi\Lz$ decays, while the decay time is modelled with the exponential distribution of the
corresponding signal mode. A simultaneous fit to both data samples is performed in order to constrain
the yield of the cross-feed and the resulting change in lifetime of $-0.3\fs$ and $+1.1\fs$ for \Bd and
$\Lb$, respectively, is assigned as a systematic uncertainty.

Another potential source of background is the incorrect association of signal $b$ hadrons to their
PV, which results in an erroneous reconstruction of the decay time. 
Since the fitting procedure ignores this contribution, a systematic 
uncertainty is evaluated by repeating the fit after including in the background model
a component describing the incorrectly associated candidates. 
The background distribution is determined by calculating the
decay time for each $\BuToJPsiK$ decay with respect to 
a randomly chosen PV from the previous selected event.
In studies of simulated events the fraction of this background is less than $0.1\%$. 
Repeating the fit with a $1\%$ contribution results in the lifetime changing by
$0.1\fs$ and, therefore, no systematic uncertainty is assigned.

The measurement of the effective lifetime in the \BsToJPsiPhi channel is integrated over the angular
distributions of the final-state particles and is, in the case of uniform angular efficiency, insensitive to the different
polarisations of the final state~\cite{LHCb-PAPER-2013-002}. To check if the angular 
acceptance introduced by the detector geometry and event selection can affect the measured lifetime, 
the events are weighted by the inverse of the angular efficiency determined in
Ref.~\cite{LHCb-PAPER-2013-002}.
Repeating the fit with the weighted dataset leads to a shift of the lifetime of $-1.0\fs$, the same as is
observed in simulation. The final result is corrected by this shift, which is also assigned as a systematic uncertainty.
The \Bs effective lifetime could also be biased due to a small \CP-odd S-wave component from
$\Bs\to\jpsi\Kp\Km$ decays that is ignored in the fit.
For the $m(\Kp\Km)$ mass range used here (Sec.~\ref{sec:selection}), 
Ref.~\cite{LHCb-PAPER-2012-040} indicates that the S-wave contribution is $1.1\%$. The effect of
ignoring such a component is evaluated by repeating the fit on simulated experiments with an additional $1\%$
\CP-odd component. A change in the lifetime of $-1.2\fs$
is observed, which is used to correct the final lifetime and is also taken as a systematic uncertainty. 
Finally, as described in Sec.~\ref{sec:selection}, only events with a decay time larger than $0.3\ps$ are
considered in the nominal fit. 
This offset leads to a different relative contribution of the heavy and
light mass eigenstates such that the lifetime extracted from the
exponential fit does not correspond to the effective lifetime defined in
Eq.~(\ref{eqn:single}). A correction of $-0.3\fs$ is applied to account for this effect.

The presence of a production asymmetry between \Bd and $\overline{\B}^0$ mesons could
bias the measured \BdToJPsiKS effective lifetime, and therefore $\Delta\Gamma_d/\Gamma_d$,
by adding additional terms in Eq.~(\ref{eqn:single}). 
The production asymmetry is measured to be
\mbox{$A_{\rm P}(\Bd) = (0.6\pm0.9)\%$}~\cite{LHCb-PAPER-2013-040},
the uncertainty of which is used to estimate a corresponding systematic
uncertainty on the \BdToJPsiKS lifetime of $1.1\fs$. No uncertainty is assigned to the \BdToJPsiKst
lifetime since this decay mode is flavour-specific\footnote{Flavour-specific 
means that the final state is only accessible via the decay of a $B^0_{(s)}$ meson 
and accessible by a meson originally produced as a $\overline{B}^0_{(s)}$
only via oscillation.} and the production asymmetry cancels in the untagged
decay rate.
For the \Bs system, the rapid oscillations, due to the large value of
$\Delta m_s = 17.768\pm0.024\invps$~\cite{LHCb-PAPER-2013-006},
reduce the effect of a production asymmetry, reported as
$A_{\rm P}(\Bs) = (7\pm5)\%$ in Ref.~\cite{LHCb-PAPER-2013-040},
to a negligible level.
Hence, no corresponding systematic uncertainty is assigned.

There is a $0.02\%$ relative uncertainty on the lifetime measurements due to
the uncertainty on the length scale of LHCb~\cite{LHCb-PAPER-2013-006}, which is
mainly determined by the VELO modules $z$ positions. These are evaluated
by a survey, having an accuracy of $0.1\mm$ over the full length of the VELO ($1000\mm$),
and refined through a track-based alignment. The alignment procedure is more precise
for the first track hits, that are less affected by multiple scattering and whose 
distribution of $z$ positions have an RMS of $100\mm$.
In this region, the differences between the module positions
obtained from the survey and track-based alignment are within $0.02\mm$, 
which is taken as systematic uncertainty.
The systematic uncertainty related to the momentum
scale calibration affects both the $b$ hadron candidate mass and momentum and, therefore,
cancels when computing the decay time.

The systematic uncertainty related to the choice of $45\fs$ for the width of the 
decay-time resolution function ($65\fs$ in the case of \BdToJPsiKS and \LbToJPsiL)
is evaluated by changing the width by $\pm 15\fs$ and repeating the fit. 
This change in width is larger than the estimated uncertainty on the resolution
and leads to a negligible change in the fit results. Consequently, no systematic 
uncertainty is assigned. Furthermore, to test the 
sensitivity of the lifetimes to potential mismodelling of the long tails in the resolution, the 
resolution model is changed from a single Gaussian function to a sum of two or three Gaussian
functions with parameters fixed from simulation. Repeating the fit with the new
resolution model causes no significant change to the lifetimes and no systematic uncertainty is assigned. 
The lifetimes are insensitive to the offset, $\Delta$, in the resolution model. 

Several consistency checks are performed to study the stability of the lifetimes,
by comparing the results obtained using different
subsets of the data in terms of magnet polarity, data taking period, $b$-hadron and track kinematic variables,
number of PVs in the event and track multiplicity. 
In all cases, no trend is observed and all lifetimes are compatible with the nominal results.

\begin{table}[t]
\caption{\small Statistical and systematic uncertainties (in femtoseconds) for the
values of the $b$-hadron lifetimes.
The total systematic uncertainty is obtained by combining the individual contributions in quadrature.\label{tab:systematics}}
\centering{
\small
\begin{tabular}{lccccc}
Source				&	$\tau_{\BuToJPsiK}$	&	$\tau_{\BdToJPsiKst}$	&	$\tau_{\BdToJPsiKS}$	&	$\tau_{\LbToJPsiL}$	& $\tau_{\BsToJPsiPhi}$	\rule[-3mm]{0mm}{0mm}
\\
\hline
Statistical uncertainty	&	3.5		&	6.1		&	12.8\, \ 		&	26.5\, \ 	&	11.4\, \ 	\\
\hline
VELO reconstruction		&	2.0		&	2.3		&	0.9		&	0.5	&	2.3	\\
Simulation sample size	&	1.7		&	2.3		&	2.9		&	3.7	&	2.4	\\
Mass-time correlation	&	1.4		&	1.8		&	2.1		&	3.0	&	0.7	\\
Trigger and selection eff.	&	1.1		&	1.2		&	2.0		&	2.0	&	2.5	\\
Background modelling	&	0.1		&	0.2		&	2.2		&	2.1	&	0.4	\\
Mass modelling			&	0.1		&	0.2		&	0.4		&	0.2	&	0.5	\\
Peaking background 	&	--		&	--		&	0.3		&	1.1	&	0.4	\\
Effective lifetime bias	&	--		&	--		&	--		&	--	&	1.6	\\
\Bd production asym.	&	--		&	--
		&	1.1		&	--	&	--	\\
LHCb length scale		&	0.4		&	0.3		&	0.3		&	0.3	&	0.3	\\
\hline
Total systematic		&	3.2		&	3.9		&	4.9		&	5.7	&	4.6	\\
\hline
\end{tabular}
}
\end{table}

\begin{table}[t]
\caption{\small Statistical and systematic uncertainties (in units of $10^{-3}$) for the lifetime ratios and
$\Delta\Gamma_d/\Gamma_d$. For brevity, $\tau_{\Bd}$ $(\tau_{\overline{B}^0})$
corresponds to the measurement of $\tau_{\BdToJPsiKst}$ $(\tau_{\overline{B}^0\to\jpsi\overline{K}^{*0}})$.
The total systematic uncertainty is obtained by combining the individual contributions in quadrature.\label{tab:systematics_ratios}}
\centering{
\small
\begin{tabular}{lccccccc}
Source				&	$\tau_{\Bu}/\tau_{\Bd}$	&	$\tau_{\Bs}/\tau_{\Bd}$	&	$\tau_{\Lb}/\tau_{\Bd}$	&	$\tau_{\B^+}/\tau_{\B^-}$	& $\tau_{\Lb}/\tau_{\Lbbar}$	&	$\tau_{\Bd}/\tau_{\overline{B}^0}$	&	$\Delta\Gamma_d/\Gamma_d$
\rule[-3mm]{0mm}{0mm}
\\
\hline
Statistical uncertainty	&	5.0		&	8.5		&	18.0\, \ 		&	4.0	&	35.0	&	8.0	&	25.0\, \ \\
\hline
VELO reconstruction		&	1.6		&	1.7		&	1.1		&	--	&	--	&	--	&	4.1\\
Simulation sample size	&	2.0		&	2.2		&	2.8		&	2.1	&	5.3	&	3.0	&	6.3\\
Mass-time correlation	&	1.6		&	1.2		&	2.3		&	--	&	--	&	--	&	4.7\\
Trigger and selection eff.		&	1.1		&	1.8		&	1.5		&	--	&	--	&	--	&	4.0\\
Background modelling	&	0.3		&	0.1		&	1.5		&	0.2	&	3.0	&	1.4	&	3.8\\
Mass modelling			&	0.2		&	0.4		&	0.2		&	0.1	&	0.2	&	0.2	&	0.8\\
Peaking background 	&	--		&	0.3		&	0.7		&	--	&	--	&	--	&	0.5\\
Effective lifetime bias	&	--		&	1.0		&	--		&	--	&	--	&	--	&	--\\
\Bd production asym.	&	--		&	--		&	--		&	--	&	--	&	8.5	&	1.9\\
\hline
Total systematic		&	3.2		&	3.7		&	4.4		&	2.1	&	6.1	&	9.1	&	10.7\, \ \\
\hline
\end{tabular}
}
\end{table}

The majority of the systematic uncertainties described above can be propagated to the 
lifetime ratio measurements in Table~\ref{tab:ratio_results}. However, some of the uncertainties
are correlated between the individual lifetimes and cancel in the ratio. For the first
set of ratios and for $\Delta\Gamma_d/\Gamma_d$, the 
systematic uncertainty from the VELO-reconstruction efficiency weights and the LHCb
length scale are considered as fully correlated.
For the second set of ratios, other systematic uncertainties, as indicated in
Table~\ref{tab:systematics_ratios}, cancel, since 
the ratio is formed from lifetimes measured using the same decay mode.
In contrast to the situation for the measurement of the \Bd lifetime in the \BdToJPsiKst mode, 
the \Bd production asymmetry does lead to a systematic uncertainty
on the measurement of
$\tau_{\BdToJPsiKst}/\tau_{\overline{B}^0\to\jpsi\overline{K}^{*0}}$ since
terms like $A_{\rm P}\cos(\Delta m_d t)$ do not cancel in the decay rates
describing the decays of \Bd and $\overline{B}^0$ mesons to 
$\jpsi K^{*0}$ and $\jpsi\overline{K}^{*0}$ final states. 
The effect of candidates where the
flavour, via the particle identification of the decay products, has not been correctly
assigned is investigated and found to be negligible.

%% file: paper-results.tex
\section{Results and conclusions}
\label{sec:results}

The measured $b$-hadron lifetimes are reported in Table~\ref{tab:sfit_fit_results}.
All results are compatible
with existing world averages~\cite{HFAG}. The reported $\tau_{\LbToJPsiL}$ is
smaller by approximately $2\sigma$ than a previous measurements from \lhcb~\cite{LHCb-PAPER-2013-032}.  
With the exception of the \LbToJPsiL channel, these are the single most precise measurements of the $b$-hadron lifetimes.
The \Bs meson effective lifetime is measured using the same data set as
used in Ref.~\cite{LHCb-PAPER-2013-002} for the measurement of the \Bs meson
mixing parameters and polarisation amplitudes in \BsToJPsiPhi decays. 
The \Bs meson effective lifetime computed from these quantities is compatible with the 
lifetime reported in this paper and a combination of the two results is, therefore, inappropriate. 

\begin{table}[t]
\caption{\small Fit results for the \Bu, \Bd, \Bs mesons and $\Lb$ baryon lifetimes. The first uncertainty is
statistical and the second is systematic.}
\centerline{
\begin{tabular}{lc}
	Lifetime		&	Value [\ps]\\
	\hline
	$\tau_{\BuToJPsiK}$	  &	1.637 $\pm$ 0.004 $\pm$ 0.003 \\
	$\tau_{\BdToJPsiKst}$&	1.524 $\pm$ 0.006 $\pm$ 0.004 \\
	$\tau_{\BdToJPsiKS}	$&	1.499 $\pm$ 0.013 $\pm$ 0.005 \\
	$\tau_{\LbToJPsiL}$	  &	1.415 $\pm$ 0.027 $\pm$ 0.006 \\
	$\tau_{\BsToJPsiPhi}	$&	1.480 $\pm$ 0.011 $\pm$ 0.005 \\
	\hline
\end{tabular}
}

\label{tab:sfit_fit_results}
\end{table}

Table~\ref{tab:ratio_results} reports the ratios of the \Bu, \Bs and $\Lb$ lifetimes
to the \Bd lifetime measured in the flavour-specific \BdToJPsiKst channel. 
This decay mode provides a better normalisation than the \BdToJPsiKS channel
due to the lower statistical uncertainty on the \Bd meson lifetime
and due to the fact that the \BdToJPsiKst lifetime only depends
quadratically on $\Delta\Gamma_d/\Gamma_d$, as shown in Eq.~(\ref{eqn:Bd_eff_Kst}).
The statistical and systematic uncertainties from the absolute lifetime measurements are propagated
to the ratios, taking into account the correlations between the systematic uncertainties.
All ratios are consistent with SM
predictions~\cite{Beneke:1996gn, Keum:1998fd, Beneke:2002rj, Tarantino:2003qw, Franco:2002fc, Gabbiani:2003pq, Gabbiani:2004tp,Lenz:2011ti}
and with previous measurements~\cite{HFAG}.
Furthermore, the ratios $\tau_{B^+}/\tau_{B^-}$, $\tau_{\Lb}/\tau_{\Lbbar}$ and
$\tau_{\BdToJPsiKst}/\tau_{\overline{B}^0\to\jpsi\overline{K}^{*0}}$
are reported. Measuring any of these different from unity would indicate a violation of 
\CPT invariance or, for
\mbox{\BdToJPsiKst} decays, could also indicate that $\Delta\Gamma_d$ is 
non-zero and \BdToJPsiKst is not $100\%$ flavour-specific.
No deviation from unity of these ratios is observed. 

\begin{table}[t]
\caption{\small Lifetime ratios for the \Bu, \Bd, \Bs mesons and $\Lb$ baryon. The first uncertainty is
statistical and the second is systematic.}
\centerline{
\begin{tabular}{lc}
	Ratio			&	Value\\
	\hline
	$\tau_{\Bu}/\tau_{\BdToJPsiKst}$		&	1.074 $\pm$ 0.005 $\pm$ 0.003 \\
	$\tau_{\Bs}/\tau_{\BdToJPsiKst}$	&	0.971 $\pm$ 0.009 $\pm$ 0.004 \\
	$\tau_{\Lb}/\tau_{\BdToJPsiKst}$	&	0.929 $\pm$ 0.018 $\pm$ 0.004 \\\hline
	$\tau_{\Bu}/\tau_{B^-}$		&	1.002 $\pm$ 0.004 $\pm$ 0.002 \\
	$\tau_{\Lb}/\tau_{\Lbbar}$	&	0.940 $\pm$ 0.035 $\pm$ 0.006 \\
	$\tau_{\BdToJPsiKst}/\tau_{\overline{B}^0\to\jpsi\overline{K}^{*0}}$		&	1.000 $\pm$ 0.008 $\pm$ 0.009 \\
	\hline
\end{tabular}
}
\label{tab:ratio_results}
\end{table}

The effective lifetimes of \mbox{\BdToJPsiKst} and \BdToJPsiKS decays are used to
measure $\Delta\Gamma_d/\Gamma_d$. 
Flavour-specific final states such as \BdToJPsiKst have
\mbox{$\mathcal{A}_{\Delta \Gamma_d}^{\BdToJPsiKst}=0$}, while $\mathcal{A}_{\Delta \Gamma_d}^{\BdToJPsiKS} = \cos (2\beta)$ to a good approximation in the SM,
where \mbox{$\beta \equiv \arg\left[-(V_{cd}V^*_{cb})/(V_{td}V^*_{tb})\right]$}
is one of the CKM unitarity triangle angles.
Hence, the two effective lifetimes can be expressed as
\begin{alignat}{2}
\label{eqn:Bd_eff_Kst}
\tau_{\Bd \rightarrow J/\psi K^{*0}} &= \frac{1}{\Gamma_d}\frac{1}{1-y_d^2} \left( 1+ y_d^2 \right),\\
\label{eqn:Bd_eff_KS}
\tau_{\Bd \rightarrow J/\psi K^0_S} &= \frac{1}{\Gamma_d}\frac{1}{1-y_d^2} \left( \frac{1+2 \cos(2\beta) y_d + y_d^2}{1+\cos(2\beta) y_d} \right).
\end{alignat}
Using the effective lifetimes reported in Table~\ref{tab:sfit_fit_results} and
$\beta = (21.5^{+0.8}_{-0.7})^{\circ}$~\cite{HFAG},
a fit of $\Delta\Gamma_d$ and $\Gamma_d$ to the expressions in
Eq.~(\ref{eqn:Bd_eff_Kst}) and Eq.~(\ref{eqn:Bd_eff_KS}) leads to
\begin{align} 
\Gamma_d &= \phantom{+}0.656 \pm 0.003  \pm 0.002\invps ,\\
\Delta \Gamma_d &= -0.029 \pm 0.016 \pm 0.007\invps ,
\end{align}
where the first uncertainty is statistical and the second is systematic. 
The correlation coefficient between $\Delta\Gamma_d$ and $\Gamma_d$ 
is $0.43$ when including statistical and systematic uncertainties.
The combination gives
\begin{equation}
\frac{\Delta \Gamma_d}{\Gamma_d} = -0.044 \pm 0.025  \pm 0.011,
\end{equation}
consistent with the SM expectation~\cite{Lenz:2006hd,Lenz:2011ti} and
the current world-average value~\cite{HFAG}.

%% file: acknowledgements.tex
\section*{Acknowledgements}
\noindent We express our gratitude to our colleagues in the CERN
accelerator departments for the excellent performance of the LHC. We
thank the technical and administrative staff at the LHCb
institutes. We acknowledge support from CERN and from the national
agencies: CAPES, CNPq, FAPERJ and FINEP (Brazil); NSFC (China);
CNRS/IN2P3 and Region Auvergne (France); BMBF, DFG, HGF and MPG
(Germany); SFI (Ireland); INFN (Italy); FOM and NWO (The Netherlands);
SCSR (Poland); MEN/IFA (Romania); MinES, Rosatom, RFBR and NRC
``Kurchatov Institute'' (Russia); MinECo, XuntaGal and GENCAT (Spain);
SNSF and SER (Switzerland); NAS Ukraine (Ukraine); STFC (United
Kingdom); NSF (USA). We also acknowledge the support received from the
ERC under FP7. The Tier1 computing centres are supported by IN2P3
(France), KIT and BMBF (Germany), INFN (Italy), NWO and SURF (The
Netherlands), PIC (Spain), GridPP (United Kingdom).
We are indebted to the communities behind the multiple open source software packages we depend on.
We are also thankful for the computing resources and the access to software R\&D tools provided by Yandex LLC (Russia).